\documentclass[%
 reprint,
 amsmath,amssymb,
 aps, onecolumn
]{revtex4-1}

\usepackage{printlen}
\usepackage{color,soul}
 
\usepackage{xcolor}
\usepackage{graphicx}
\usepackage{dcolumn}
\usepackage{bm}
\usepackage{hyperref}
\usepackage[mathlines]{lineno}%
\usepackage{subfigure}
\usepackage[normalem]{ulem}

\newcommand*{\affaddr}[1]{#1} 
\newcommand*{\affmark}[1][*]{\textsuperscript{#1}}
\begin{document}

\preprint{APS/123-QED}

\date{\today}

\title{How Language, Culture, and Geography shape Online Dialogue: Insights from Koo}
\author{Amin Mekacher\affmark[1], Max Falkenberg\affmark[1] and Andrea Baronchelli\affmark[1,2,*]
\begin{center}
\affaddr{\affmark[1]{ \textit{City University of London, Department of Mathematics, London EC1V 0HB, (UK)}}} \\
\affaddr{\affmark[2]{ \textit{The Alan Turing Institute, British Library, London NW1 2DB, (UK)}}} \\
\affaddr{\affmark[*]{Corresponding author: abaronchelli@turing.ac.uk}}
\end{center}
}
\begin{abstract}
\vspace{0.5cm}
\textbf{Koo is a microblogging platform based in India launched in 2020 with the explicit aim of catering to non-Western communities in their vernacular languages. With a near-complete dataset totalling over 71M posts and 399M user interactions, we show how Koo has attracted users from several countries including India, Nigeria and Brazil, but with variable levels of sustained user engagement. We highlight how Koo's interaction network has been shaped by multiple country-specific migrations and displays strong divides between linguistic and cultural communities, for instance, with English-speaking communities from India and Nigeria largely isolated from one another. Finally, we analyse the content shared by each linguistic community and identify cultural patterns that promote similar discourses across language groups. Our study raises the prospect that a multilingual and politically diverse platform like Koo may be able to cultivate vernacular communities that have, historically, not been prioritised by US-based social media platforms.}
\end{abstract}
\maketitle

\section*{Introduction}

The social media ecosystem, which has historically been Western-focused \cite{wike2022_democracy}, has evolved significantly in recent years, with a rapidly growing number of the active users from non-Western countries and the Global South \cite{poushter2018_social}. Despite this shift in demographics, major social media platforms such as Twitter and Facebook lack adequate support for many major vernacular languages \cite{mandavia2021_vernacular} (e.g., in South-East Asia \cite{leong2022_asia}), with platforms continuing to prioritise Western audiences. For instance, investments into English-language content moderation on Twitter, Facebook and Instagram still heavily outstrip investments into moderation tools for other language's \cite{milmo2021_facebook}, indicating that platforms are less equipped to tackle, and are (arguably) less concerned about, harmful content posted in languages other than English. This is despite documented evidence that social media can play a key role in election campaigns, and the fact that India, the World's largest democracy with upcoming elections in 2024, has become the largest market in terms of users for many leading social platforms including Instagram \cite{statista2023_instagram}, Youtube \cite{mbw2022_youtube} and Facebook \cite{vengattil2022_facebook}. 

This persistent failure to support many communities beyond the West is the \textit{raison d'etre} of Koo, a microblogging platform launched in Bangalore, India, in early 2020. Koo aims to champion a ``language-first'' approach, where each user is able to express themselves in their native language when connecting with their peers \cite{radhakrishna2022_koo}. By upscaling their auto-translate tool, Koo aims to offer an inclusive experience to speakers of less widely utilized languages, a feature not prioritised on the dominant US-based platforms. Indicative of this, Koo currently supports 20 of India's 22 official languages, whereas Twitter only supports 5 \cite{twitter2024_languages}. By appealing to political leaders across countries, mostly from India, Nigeria and Brazil, including some who have criticised US-based social media platforms, Koo has managed to attract a geographically diverse user-base, becoming the second largest microblogging platform globally after Twitter \cite{singh2022_koo, inamdar2022_koo}. As such, it occupies an influential position in the social media ecosystem and offers a unique opportunity to study the role of language on social media. 

This paper follows our recent release of the Koo dataset \cite{mekacher2024_zenodo}, and an accompanying paper in which we discuss the platform's sustainability (relative to other platforms), its use in political debates, and how news is shared on the platform \cite{mekacher2024_koo}. Although often defined as an alt-tech platform, Koo has attracted a more international user base than US-based alt-tech platforms \cite{mekacher_2022voat, ali2021_deplatforming, papasavva2023_poal} leading to a more diverse community. Here, we extend this work by focusing explicitly on the role that language plays in shaping the structure of a non-Western social platform. 

Previous social media studies have considered the role of language in shaping online interactions, but not in the context of India. Researchers have studied linguistic trends on Twitter and found that English-speaking posts were dominant on the platform when it mostly attracted users from Western nations \cite{Hong2021_lang, Mocanu2013_lang}, whereas the usage of English became less prevalent when considering non-Western countries \cite{Krishnamurthy2008_twitter, Java2007_twitter}. Moreover, following the growth of communities outside English-speaking nations, different linguistic communities were shown to interact differently with a social platform's features, leading to distinct social structures \cite{weerkamp2011_lang}. However, despite the subsequent globalisation of social media, the formation of dyadic ties on Twitter was found to be strongly correlated with the linguistic background of a user, even between different English-speaking countries \cite{Takhteyev2012_language, falkenberg2023_polarization}. Studies have also considered the influence of bilingual social media users on interaction networks, and whether users post in languages other than a platforms dominant language \cite{Kim2014_multilingual, Androutsopoulos2007_language}. Language diversity was further quantified by using geolocation data to map the language diversity in the Greater Manchester \cite{bailey2013_language}. More recently, studies have compared literacy levels across regions on Facebook \cite{Lin2023_facebook}. Our research contributes to this literature by studying an online platform striving to host vernacular languages, in the hope to harbour a sustainable multi-lingual community.

With 22 nationally-recognised languages, and over 100 languages with more than 10,000 native speakers, India is a unique case study to assess the impact of linguistic pluralism on user-to-user interactions online. India's linguistic history has been the focus of several studies, looking into the national linguistic landscape \cite{apte1976_multilingualism}, the patterns of communication \cite{mallikarjun2010_patterns} and the socio-economic ramifications of a complex linguistic environment \cite{pattanayak1990_multilingualism}, but many of these methods have not been applied in the context of social media. Moreover, the rich linguistic composition of India also allows for an analysis of language use at a national scale, before looking at linguistic communities across several countries. Similar observations are valid for Nigeria, one of the other countries where Koo was adopted by government officials, where multilingualism plays a major role in social interactions in several areas \cite{kari2002_multilingualism}, with about 500 vernacular languages spoken across the country \cite{moses2012_nigeria}.

In the remainder of this paper we first assess the impact of the various collective migrations to Koo which shaped the platform into a multilingual venue. We then study how political incentives to migrate to Koo led to different degrees of user engagement over time. Afterwards, we look at the topology of the interaction network, while assessing the impact of the language barrier to foster cross-cultural communities. Finally, we look at user mobility across linguistic landscapes and how this relates to the richness of a community's online conversation, as well as the shared discourse between language pairs. Our findings suggest that linguistic and cultural factors are instrumental in bridging communities on Koo, with few interactions taking place across communities with different linguistic backgrounds.

\section*{Results}

\subsection*{Platform migration and user retention}

\begin{figure}[h!]
    \centering
    \includegraphics[scale=0.8]{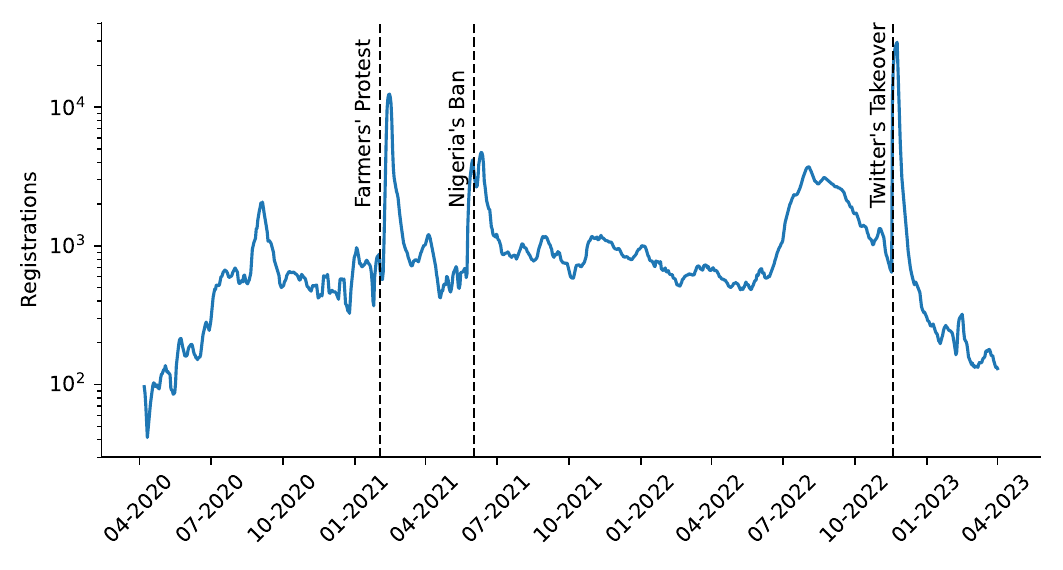}
    \caption{\textbf{Daily number of registrations on Koo, and the impact of collective migration.} 7-day moving average of the daily number of registrations on Koo, from the beginning of 2020 to early 2023. The dashed lines indicate, in order:  the migration of BJP politicians and their supporters following the Indian Farmers' Protest in February 2021; the migration of the Nigerian government after Twitter was banned in the country in June 2021; the Brazilian community joining Koo in November 2022 after Elon Musk purchased Twitter.}
    \label{fig:migration}
\end{figure}

We begin by examining Koo's popularity over time in the online ecosystem. Figure \ref{fig:migration} shows the daily number of registrations on Koo, from the launch of the platform in 2020 until early 2023. Major political and social events, which had an impact on Koo's outreach, are marked as dashed lines. 

The first significant peak in registrations can be seen in February 2021. India was in the midst of the farmers' protest, a popular movement against a new set of laws adopted by the Parliament of India in September 2020. These events triggered a conflict between the Indian government and Twitter after BJP (the ruling political party) officials pressured Twitter to ban accounts linked to the popular movement \cite{ellis2023_twitter}. Members of the government and BJP supporters in India subsequently signed up to Koo, as Twitter did not comply with their requests, and invited their community to follow suit \cite{gerry2021_farmer}. As seen on the figure, the political movement managed to increase Koo's user base substantially \cite{nilesh2021_koo}. The platform's willingness to comply with content take down orders issued by the government made Koo more attractive to BJP politicians, thereby cementing the dominance of BJP narratives on the platform \cite{gerry2021_farmer}.

The second burst in registrations dates from June 2021, when Nigerian then-President Muhammadu Buhari banned Twitter from the country and registered on Koo with members of his government \cite{ayomide2022_nigeria}, after Buhari's tweets were deleted for inciting violence against his political opponents \cite{princewill2021_nigeria}. Koo experienced a wide adoption from government officials in Nigeria, prompting the platform to hire vernacular speakers for content moderation purposes in Nigeria \cite{abubakar2021_nigeria}. The government was also followed by a large number of Nigerian users who subsequently signed up to Koo, leading to an uptick in registrations. However, Koo had little success in attracting Nigerian celebrities or influencers, unlike in India where the platforms gained support from Bollywood actors and prominent cricket players \cite{phartiyal2021_india}. Previous research has shown that celebrities' endorsement can catalyse a massive migration towards alt-tech platforms \cite{mekacher2023_gettr}. 

The last major peak in registrations took place in November 2022, shortly after Twitter was purchased by Elon Musk. Felipe Neto, a Brazilian influencer with over 16 million followers on Twitter, advertised his migration to Koo on Twitter, which led to his followers signing up to the platform as well \cite{gonzalez2022_brazil}. This collective movement was strengthened when Brazilian President Lula also registered on Koo \cite{regina2022_lula}. In total, Koo's user base grew substantially in Brazil, with the Koo app downloaded over 1 million times in the space of 48 hours \cite{reuben2022_brazil}.

These three events have shaped the major linguistic communities on Koo. A breakdown of the registration numbers broken down by language is also provided in the SI, showcasing the influx of Hindi, Nigerian English, and Portuguese users. 

To assess the success of linguistic migrations, we measure user retention, i.e. how many users within a cohort are still active after a given number of days. Throughout our analysis, we match each user with the language they used the most when posting and commenting on the platform.  

\begin{figure}[h!]
    \centering
    \includegraphics[scale=0.8]{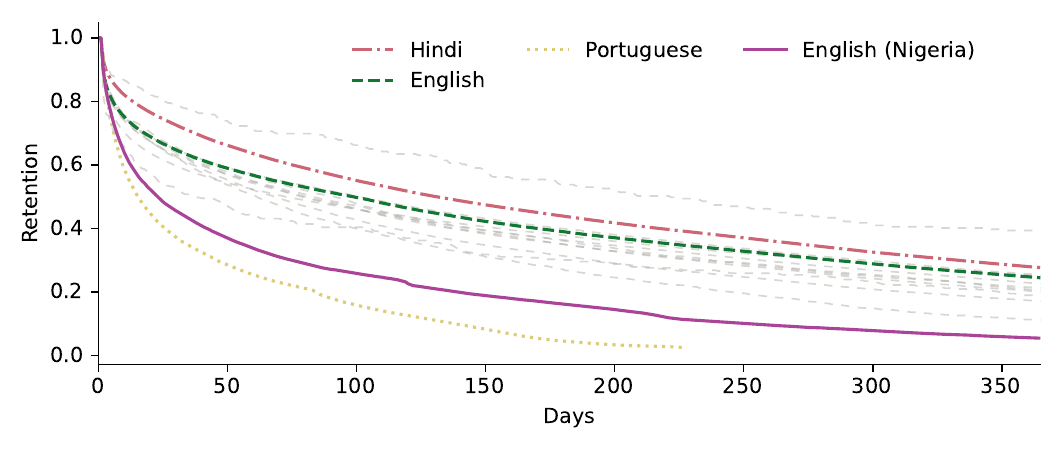}
    \caption{\textbf{Heterogeneous user retention for various linguistic communities.} Kaplan-Meier survival curves for the main linguistic communities on Koo, showing the fraction of users who remained active after a given number of days. For each user, we define ``day zero'' as being their registration date on Koo. Other linguistic communities are displayed in grey. The retention curve is displayed until the day that fewer than 1\% of users from a linguistic community remain active.}
    \label{fig:survival}
\end{figure}

Figure \ref{fig:survival} shows the Kaplan-Meier estimator, a tool used to visualise the retention curve of a population over time, computed for each linguistic community on Koo. Given a linguistic community, the Kaplan-Meier estimator indicates how many users were still active on Koo, a given number of days after they registered on the platform. The figure indicates that both Brazilian and Nigerian communities have a much lower retention than other linguistic communities on Koo, with 50\% of the cohort becoming inactive within 16 days and 23 days of signing up to the platform, respectively. In contrast, it took 131 days to reach the same level of user retention when considering Hindi-speaking users, thus highlighting a strong difference in user engagement across linguistic communities and countries. The smaller linguistic clusters also display a higher survival rate than the Brazilian and Nigerian users, which suggest that the sustainability of a community does not only depend on its population size.

The lack of long-term adoption in the Nigerian community can be explained by a popular resistance to the Twitter ban that was instated by Buhari's government in June 2021. Nigerian users managed to bypass the ban shortly after it was instated, with VPN usage becoming more common nation-wide \cite{ayomide2022_nigeria}. Moreover, Koo did not receive sustained support from the Nigerian government. Muhammadu Buhari lifted the ban on Twitter in January 2022, after the platform and his government settled on an agreement \cite{akinwotu2022_nigeria}. Buhari stopped being active on Koo shortly afterwards, as did most of the government members who joined the platform. Koo's monthly active users in Nigeria fell by over 90\%, suggesting that the platform failed to establish a foothold as sustainable as their popularity in India \cite{dosunmu2023_nigeria}. 

In the case of Brazil, the migration was not, primarily, triggered by political motivations in the same way as for India and Nigeria, but rather by a linguistic pun involving the word ``koo'' in Portuguese, although Brazilian President Lula did join the platform during its initial growth-phase. However, in general, the Brazilian community and Brazilian celebrities did not stay as engaged on Koo, when compared to celebrities from India \cite{livemint2023_koo}. Both Felipe Neto and Lula, who were the main drivers of the Brazilian migration on Koo, stayed active on Twitter and their followers can therefore still follow their feed without requiring access to Koo. Previous research has shown that users are less active on alt-tech platforms if they can still reach out to their followers via a mainstream outlet \cite{mekacher2023_gettr}. As of November 2023, Lula is still sporadically active on Koo, whereas Felipe Neto's last interaction dates to May 2023.

These results suggest that collective migrations to an alternative social platform can have mixed levels of success, depending on the motivations triggering the migration and the degree of approval it garners across the community. The Indian migration exemplifies the birth of a sustainable community on Koo, as it was led by government officials and garnered support from both national celebrities and BJP supporters. The Nigerian government followed a different pattern, where the low support for the Twitter ban outside of Buhari's supporters led to a short-lived retention for most Nigerian users who signed up on Koo. In the same fashion, the Brazilian community shows a low level of user retention, which can be explained by a lack of social incentives to shift the political discourse to an alternative platform, and away from the dominant US-based platforms.

\subsection*{Language-use in the Koo interaction network}

We now focus on user interactions on Koo and the landscape that emerges on a platform where many languages and cultures coexist, following individual migration decisions. 

\begin{figure}[h!]
    \centering
    \includegraphics[scale=0.85]{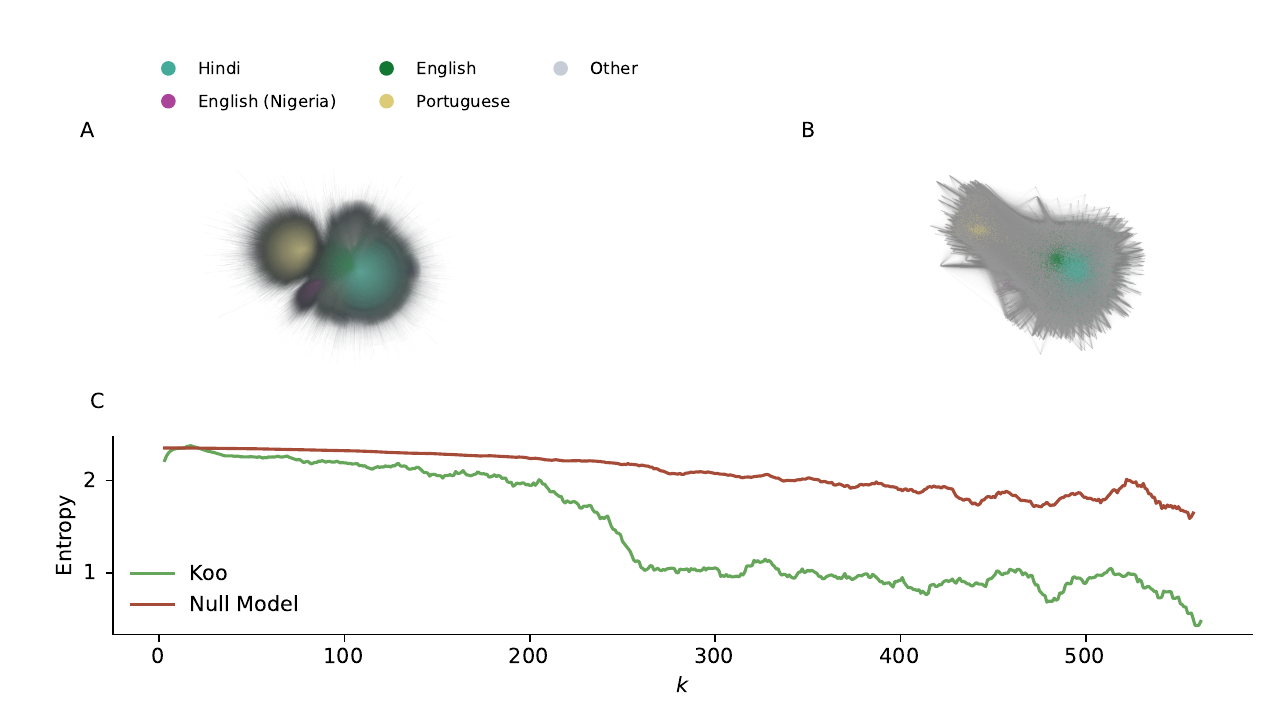}
    \caption{\textbf{The Koo interaction network and the impact of linguistic homophily on the network's structure.} Each node represents a user, and two nodes are connected if one of the users interacted with the other user's content. Users are coloured according to their modal language on the platform. The main linguistic communities are the Hindi-speaking users (blue), English-speaking users (green), Nigerian users (purple) and Portuguese-speaking users (yellow). The layout is generated by using a force-directed graph drawing method. A) The total interaction network. B) The $k$-core of the interaction network with $k = 150$. C) The Shannon entropy of the modal language of the nodes belonging to the k-core of the graph, with respect to the value of $k$. The entropy of the interaction network is compared to the value obtained in a null model, where we shuffle the modal language associated to each node in the network.}
    \label{fig:interaction_kcore}
\end{figure}

Figure \ref{fig:interaction_kcore}A shows an interaction network, where two users are connected if they interacted (i.e. one of the users liked, shared or commented the other's post) on Koo, with the weight of an edge proportional to the number of interactions between a pair of users; for simplicity we treat the interaction network as undirected. The network layout, generated with a force-directed graph drawing method, highlights the strong segregation between several linguistic communities: Portuguese-speaking users (yellow) mostly interact with other Portuguese-speaking users, and likewise, Hindi-speaking users (blue) mostly interact with other Hindi-speaking users. On the other hand, the English-speaking cohort (green) acts as a bridge between Hindi, Portuguese and Nigerian English speakers (purple), as well as the smaller cohorts that can be seen in the periphery of the graph.

The network indicates a strong homophilic behaviour, meaning that users will mainly engage with members of their own linguistic community. Koo's auto-translate feature, which allows users to translate any post to the language of their choice, does not seem to mitigate the language assortativity that we observe on the platform. Homophilic patterns have been observed in many online social spaces, for example when considering information diffusion dynamics \cite{choudhury2010_homophily}, community formation \cite{Dehghani2016_homophily}, and in political interaction networks \cite{falkenberg2023_polarization}. 

We can highlight this homophilic behaviour by computing the $k$-cores of the network. Given a network, its $k$-core for $k \in \mathbb{N}$ is defined as the sub-graph in which all vertices have a degree greater than or equal to $k$. $k$-core analysis offers a deeper view into the tightly connected components of a network, therefore identifying the nodes that are strongly interconnected or play an influential role in the overall network structure \cite{Algaradi2017_kcore}. Figure \ref{fig:interaction_kcore}B shows the interaction network from figure \ref{fig:interaction_kcore}A but filtered so that only the $k \geq 150$ core is shown. This threshold allows us to filter out the periphery of the network, by only keeping the 5\% most strongly connected nodes of the network.

This visualisation offers a better overview of the tightly interconnected components within the interaction network, with a strong core of English-speaking and Hindi-speaking users. Conversely, the Brazilian and Nigerian clusters are more isolated. The $k$-core also highlights that users belonging to smaller linguistic communities are scarcely included in the core of the network, with 95\% of the core users belonging to the Hindi, English, Nigerian English and Portuguese-speaking clusters. This analysis further underlines that highly connected clusters mostly involve users who speak one of the dominant languages on the platform. Looking at the English-speaking community, we note that it is principally connected to Hindi speakers in the $k$-core, highlighting the instrumental role that the English language plays in Indian political communications \cite{Schiffman1998_hindi}. On the other hand, both the Nigerian and the Brazilian communities dominantly communicate in their native language and are therefore isolated in the $k$-core. 

To quantify the connection between linguistic communities, we generate the $k$-core for all values of $k$ for which the $k$-core exists. We retrieve the modal language of each vertex included in the $k$-core and compute the Shannon entropy, to evaluate whether the $k$-core encompasses a diverse range of languages or is primarily dominated by a few major languages \cite{Kang2023_entropy, Kepinska2023_entropy}. Once we have computed the $k$-core, we retrieve the modal language of each user included in the core and calculate the entropy of the list of languages. Figure \ref{fig:interaction_kcore}C displays the resulting entropy with respect to $k$. We notice that higher values of $k$ lead to a lower entropy in the language composition of the $k$-core, suggesting that dense interactions on Koo take place mostly within homogeneous linguistic clusters.

To ensure that the lack of diversity in high-degree interactions is a characteristic of the interaction network on Koo, and not an erroneous finding, we define a null-model of the network, where the modal language is shuffled for the nodes in the network (preserving the language prevalence distribution), and the Shannon entropy is computed again for each value of $k$. Figure \ref{fig:interaction_kcore}C displays the median Shannon entropy for each value of $k$, after running the null-model 1000 times. We notice that the entropy for the null-model does not sharply decrease for higher values of $k$, which reveals that this sharp decline in language diversity within the k-core is indeed a distinctive feature of the Koo interaction network. These findings further indicate that cross-linguistic interactions on Koo are rare with respect to same-language interactions.

To measure the prevalence of a linguistic community within the k-core for any value of $k$, vertices in the network can also be defined by their coreness, i.e., the maximum value of $k$ for which they still belong to the $k$-core. In the SI, we show the distribution of the coreness of the users belonging to each linguistic community. Our analysis reveals that only Hindi- and English-speaking users are included in the highest cores. This result indicates that, despite the presence of a rich linguistic landscape on the platform, strong interaction ties on Koo are mainly driven by linguistic homophily, and that cross-linguistic interactions are rare.

\subsection*{Language homophily and multilingual activity}

The structure of the interaction network indicates that the linguistic background of a user strongly influences their interaction patterns on Koo. Previous studies have highlighted that many ties in social networks are strongly assortative, i.e., they connect individuals who share similar attributes in terms of cultural background or social status \cite{Newman2002_assortativity}. Language assortativity has been previously found to influence mating decisions \cite{Nagoshi1990_assortativity}, friendship ties among adolescents \cite{Leszczensky2019_ethnicity}, and political communication between countries using a common language \cite{falkenberg2023_polarization}. 

When considering a diverse population such as the Koo user base, a question to consider is the interaction patterns for members of minority linguistic groups. Previous studies have shown that minority groups in organisations rely on out-group interactions to be connected to the centre of the network \cite{Leonard2007_minority}, whereas belonging to an underrepresented social group leads to a stronger in-group identity in friendship networks \cite{Mehra1998_minority}. As such, we will next look at linguistic behaviours on Koo, by measuring the propensity for users to interact within their linguistic community on the platform. We will also evaluate a user's likelihood of using their modal language when interacting with their peers on Koo. 

To measure a user's adherence to their modal language relative to their propensity to use other languages, we define the commitment: given a user with N posts in their modal languages and M posts in other languages, the commitment is given by
\begin{equation}
    C = \frac{N}{N + M}.
\end{equation}

Commitment has previously been used in linguistic studies to assess the adoption of new linguistic norms, indicating that outdated norms are still persistently used by a minority of the population \cite{Amato2018_language}. Similar findings have been highlighted in ethnographic studies looking at the adoption of new spelling rules in both Spanish and English-speaking nations \cite{scragg1974_history, vivian1979_spelling}. The literature therefore suggests the potential for smaller-scale communities to survive in a linguistic setup, despite the rise of a dominant linguistic framework.

\begin{figure}[h!]
    \centering
    \includegraphics[scale=0.8]{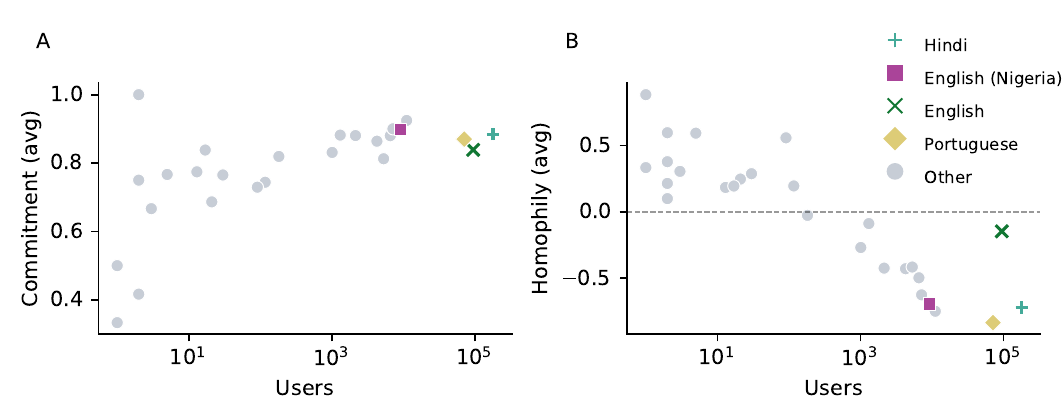}
    \caption{\textbf{EI-Homophily and language commitment and the impact of a community's size on its sustainability.} Number of users belonging to a linguistic community plotted against A) their commitment to their modal language, and B) their EI homophily index. Both metrics are averaged by the number of users for whom the language measured is their modal language. The coloured dots represent the Hindi-speaking community (blue), English (green), Portuguese (yellow) and Nigerian English (purple). The dashed line indicates an average homophily equal to 0.}
    \label{fig:commital_homophily}
\end{figure}

The average commitment of a linguistic community, plotted against its population size, is displayed on Figure \ref{fig:commital_homophily}A. We notice a strong tendency for the average commitment of a user to their modal language to increase as the population size of their community increases. Among the highlighted communities, Nigerian English speakers display the highest average commitment ($C = 0.90$), and English-speaking users also display a high level of commitment ($C = 0.84$), despite their role as a bridging community between other languages. Portuguese and Hindi speakers also have a high commitment ($C = 0.87$ and $C = 0.88$, respectively), which indicates that their communication on Koo mostly relies on their modal language, emphasising the absence of cross-language interactions. However, smaller linguistic communities display a lower commitment to their modal language. For example, Indian communities, such as Odia speakers ($C = 0.73$), and non-Indian clusters such as Spanish speakers ($C = 0.74$), are less committed to their modal language. This finding highlights the need to communicate in other languages in order to be part of a community on the platform. The attractiveness of a language has previously been modelled by its number of speakers \cite{Abrams2003_language} suggesting that smaller linguistic communities are less likely to attract new speakers, thus fuelling their need to use other languages in order to be connected to core conversations on the platform.

The high level of commitment observed for the major linguistic communities, along with the topology of the interaction network, shows that there is a strong trend for interactions on the platform to involve two users with a similar linguistic background. To measure whether social interactions on Koo mostly take place within the confines of homogeneous linguistic clusters, we use the External-Internal (EI) homophily index \cite{coleman1958_homophily}: given a node in the interaction network with E edges to their out-group (in this case, interactions with another linguistic community) and I edges with their in-group (members of the same linguistic community), their EI homophily index is given by
\begin{equation}
    EI = \frac{E - I}{E + I}.  
\end{equation}

A node which only interacts within their in-group (same language) therefore has $EI = -1$, whereas a node which only interacts with their out-group (different languages) has $EI = 1$. The EI-homophily index has previously been used to measure people's tendency to interact with their politically-aligned peers on social media \cite{Warton2022_homophily} and those sharing a similar vaccination status \cite{Are2023_vaccine}, with both studies showing an overall trend for people to cluster in homogeneous groups. A social network where the majority of interactions are intra-group links (i.e. with the EI-homophily index close to -1) is referred to as being \textit{homophilic}, whereas it is referred to as \textit{heterophilic} if the network displays several inter-group interactions (i.e. with an EI-homophily index close to 1).

The EI-homophily, averaged for each linguistic community by considering the modal language of each user, is shown in Figure \ref{fig:commital_homophily}B, and plotted against the size of each language's population. We notice that small communities have a positive homophily index, indicating that members of small linguistic clusters interact mostly with other linguistic communities. On the other hand, with the exception of the English-speaking community, larger linguistic communities are more involved in in-group interactions, leading to a negative EI-homophily index. For example, Portuguese-speaking users have an average EI-homophily index of $-0.94$, whereas it is equal to $-0.66$ for the Hindi-speaking users. This finding highlights the existence of siloed communities on Koo, where users' interactions are strongly influenced by language similarities. The strong assortativity with respect to language is further displayed in the layout of the interaction network in Figure \ref{fig:interaction_kcore}A, where we see how disjointed the Portuguese-speaking and Hindi-speaking communities are with respect to other linguistic communities. 

Looking at the average EI-homophily index for English-speaking communities in Figure \ref{fig:commital_homophily}B, we notice a stark difference between Nigerian English speakers ($EI = -0.64$) and other English speakers ($EI = -0.15$). This can also been observed in the interaction network in Figure \ref{fig:interaction_kcore}A, where Nigerian English-speaking users are disconnected from the core of the network, whereas English speakers are strongly involved in cross-language interactions, leading to a higher average EI-homophily index than for Hindi and Portuguese speakers. Thus, the English language acts as a lingua franca on Koo, allowing users from diverse linguistic backgrounds to engage in inter-community interactions. However, the contrast in homophily between Nigerian English and English speakers also highlights that language is not the only factor playing a central role in shaping the interaction network. Cultural similarities are also influential in bridging users together on a social platform. Communities can be structured around specific salient topics of conversation related to the cultural and political landscape of a country, another feature of a platform hosting diverse demographics that we investigate below. Moreover, it is likely that most of the English speakers on Koo (excluding those who speak Nigerian English) have an Indian focus, due to the ubiquity of the English language in India's public affairs \cite{Karat1972_india}.

The homophilic patterns observed in the interaction network suggest that communication across linguistic communities is rare on Koo. However, our analysis does not take into account users who communicate on the platform in more than one language, and therefore belong to more than one linguistic community on the platform. To measure the propensity for users to switch between languages, we map languages as the nodes of a network, with weighted edges representing the number of users who posted in both languages on Koo. This layout follows the principle of a global language network (GLN), which allows us to quantify indirect communications between pairs of languages by looking at the number of speakers they share \cite{ronen2014_gln}. By considering the number of modal speakers in two linguistic communities and the number of speakers they share, we use the phi coefficient to measure the association between the two languages. For two languages $i$ and $j$, their phi coefficient $\Phi_{ij}$ \cite{ronen2014_gln} is given by: 
\begin{equation}
    \Phi_{ij} = \frac{M_{ij}N - M_iM_j}{\sqrt{(M_iM_j(N - M_i)(N - M_j))}},
\end{equation}
with $N$ being the total number of users, $M_i$ and $M_j$ being the number of speakers of language $i$ and $j$ respectively, and $M_{ij}$ being the number of bilingual speakers for languages $i$ and $j$. A positive phi coefficient indicates that the number of bilingual speakers between languages $i$ and $j$ is higher than what could be expected based on their representation in our dataset, whereas a negative value indicates that the co-occurrence of both languages is underrepresented relative to the size of the communities. This metric therefore allows us to assess whether there is a stronger connection between two linguistic communities on Koo, than is expected due to chance alone.

To ensure that the link between two linguistic communities is significant, we use the \textit{t} statistic, defined as:
\begin{equation}
t_{ij} = \frac{\Phi_{ij}\sqrt{D - 2}}{\sqrt{1 - \Phi_{ij}^2}},
\end{equation}
where the degree of freedom $D$ is defined as $D = \text{min}(M_i, M_j)$. As all the linguistic communities we consider in our analysis have at least 20 modal speakers, we set $D$ = 20. By setting $p = 0.05$, we can reject the null hypothesis, i.e., the number of links between two languages in the global language network is not statistically significant, if $t_{ij} \geq 1.72$ (one-tailed $t$-test). Any significant link in the network indicates that there are significantly more bilingual speakers between two languages than expected by chance.

\begin{figure}[h!]
    \centering
    \includegraphics[scale=0.8]{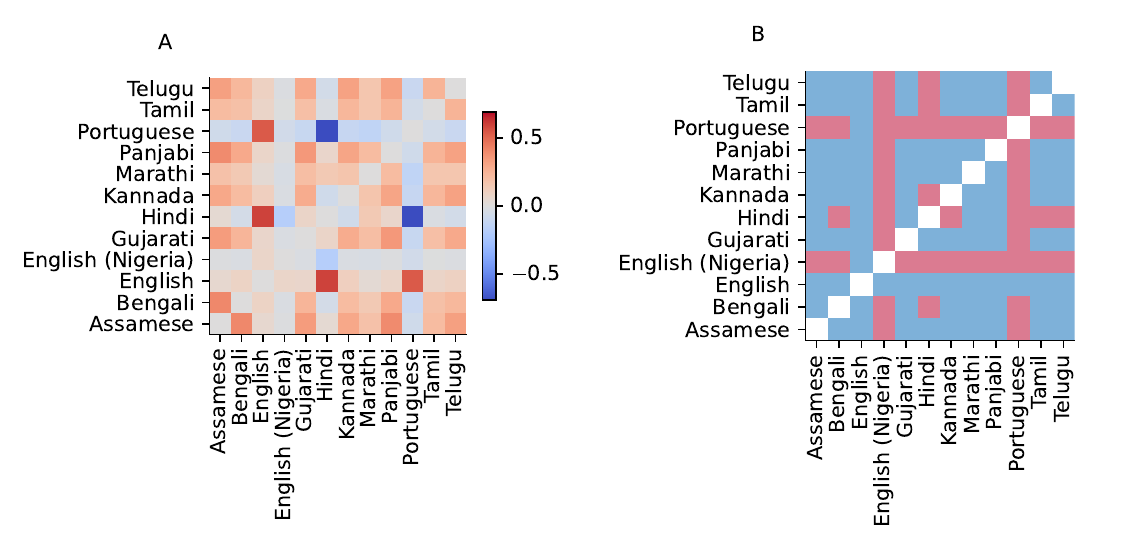}
    \caption{\textbf{Global language network and multilingual activity.} A) The correlation measured from the global language network. Two languages with a positive correlation share more connections than expected based on their respective number of speakers, and is negative otherwise. B) The t-statistic for each pair of languages in the global language network. Blue cells indicate that the link between the two languages is significant with respect to the t-statistic, whereas red cells highlight non-significant links. A link is considered significant if $p < 0.05$.}
    \label{fig:corr_t_test}
\end{figure}

Figure \ref{fig:corr_t_test}A displays the value of the phi coefficient between the main languages used on Koo. We notice that there is a positive association between the Indian languages used on Koo. Despite being strongly homophilic, the Hindi-speaking community has a positive correlation with several Indian languages, such as Gujarati and Marathi. Smaller linguistic communities in India also show a strong symbiosis in the network: languages such as Telugu, Kannada, Tamil and Assamese are strongly connected to one another, indicating that speakers of less prominent Indian languages on the platform are more likely to also communicate on Koo using another national language. Our findings are aligned with the results of the 2011 Indian language census, indicating a large number of bilingual and trilingual speakers among the smaller linguistic communities in the country \cite{orgi2011_atlas}. On the other hand, both the Nigerian English and Portuguese-speaking communities have a negative correlation with respect to every other linguistic community, English excepted.

Figure \ref{fig:corr_t_test}B shows the results of the t-statistic, with significant links in the global language network highlighted in blue. We notice that both the Brazilian and Nigerian communities share non-significant links with the smaller linguistic communities on Koo, indicating that there is not a meaningful number of users who use both Portuguese or Nigerian English and another language on Koo - with the exception of English, which is significantly correlated to all languages. These results further suggest that the Brazilian and Nigerian communities are strongly isolated when it comes to language mobility, whereas a shared cultural background enables Indian language speakers to navigate through different linguistic communities. While enhancing the access to posts written in a user's non-native language, the auto-translate feature on Koo does not appear to incentivize users to interact in languages outside their home country.

\subsection*{Discourse richness and similarity across languages}

What about content? A natural question is whether stronger ties between two linguistic communities also implies that their respective discourses are similar Moreover, some of the linguistic communities being larger than others, we hypothesise that a larger community should have a richer discourse. Studies have shown that the size of a community defines its propensity to sway the discussion topics in another community \cite{Yang2021_discourse}, and that nurturing a local discourse can allow a local community to claim their own governance \cite{Medina2009_discourse}.

To answer these questions, we use diversity measures defined in ecology, which were defined to assess the richness of an environment by looking at the presence of various species and their respective prevalence \cite{Willis2019_alpha}, as well as how often these species can be found across different environments \cite{Su2021_beta}. These methods were also previously used in linguistics research, for example to measure linguistic diversity between Canadian cities \cite{Grin2022_diversity}. For our analysis, we consider hashtags used within a linguistic community as a proxy for the discourse. Hashtags have been shown to occupy a different linguistic function than words, sharing similarities across languages \cite{mahfouz2020_hashtags}, thus allowing us to capture narratives shared by various linguistic communities on Koo. Hashtags are also a more reliable signal to measure the overlap of narratives across linguistic communities than plain text, as hashtags can be identified without the need to compare textual data from different languages. Our approach is further motivated by the presence of several low-resource languages in our corpus, which are known to be under-represented in many large language models and can therefore lead to unreliable text classification \cite{nguyen2023_llm}.

To assess the richness of the discourse within a linguistic community, we use the alpha diversity, which measures the proclivity for an environment to host various species at a local scale. In our case, we compute the alpha diversity of hashtags used by a linguistic community using the Chao1 estimator. We can also estimate the propensity for two linguistic communities to discuss similar topics by computing the beta diversity, which assesses the propensity for the species composition of two environments to be similar. Using the hashtags, we measure the beta diversity of the discourse between two linguistic communities with the Bray-Curtis dissimilarity index. 

To measure the richness of the discourse within a linguistic community, we compute the alpha diversity with the Chao1 estimator $\hat{S}_{Chao1}$, which aims at providing a lower-bound estimation of the number of unseen species, in order to assess the total number of unique hashtags used by a linguistic community from our observations. The Chao1 estimator is defined as:
\begin{equation}
    \hat{S}_{Chao1} = \begin{cases}
S + \frac{N - 1}{N} \frac{s_1^2}{2s_2} & s_2 > 0\\
S + \frac{N - 1}{N} \frac{s_1(s_1 - 1)}{2} & s_2 = 0
\end{cases},
\end{equation}
where $S$ is the total number of hashtags, $N$ is the number of unique hashtags and $s_i$ is the number of hashtags that appear at least $i$ times in the observations. This formula, however, only takes into account hashtags that appear once or twice in a linguistic community. To extend the amount of information used to estimate the richness of the conversation, we compute the improved Chao1 estimator $\hat{S}_{iChao1}$, which aims at correcting the first-order bias of $\hat{S}_{Chao1}$ by also including the number of triplets $s_3$ and quadruplets $s_4$. The improved Chao1 estimator is defined as \cite{chu2014_chao}:
\begin{equation}
    \hat{S}_{iChao1} = \hat{S}_{Chao1} + \frac{N - 3}{4N} \frac{s_3}{s_4} \cdot max\big(s_1 - \frac{N -3}{N - 1}\frac{s_2s_3}{2s_4}, 0\big).
\end{equation}

\begin{figure}[h!]
    \centering
    \includegraphics[scale=0.8]{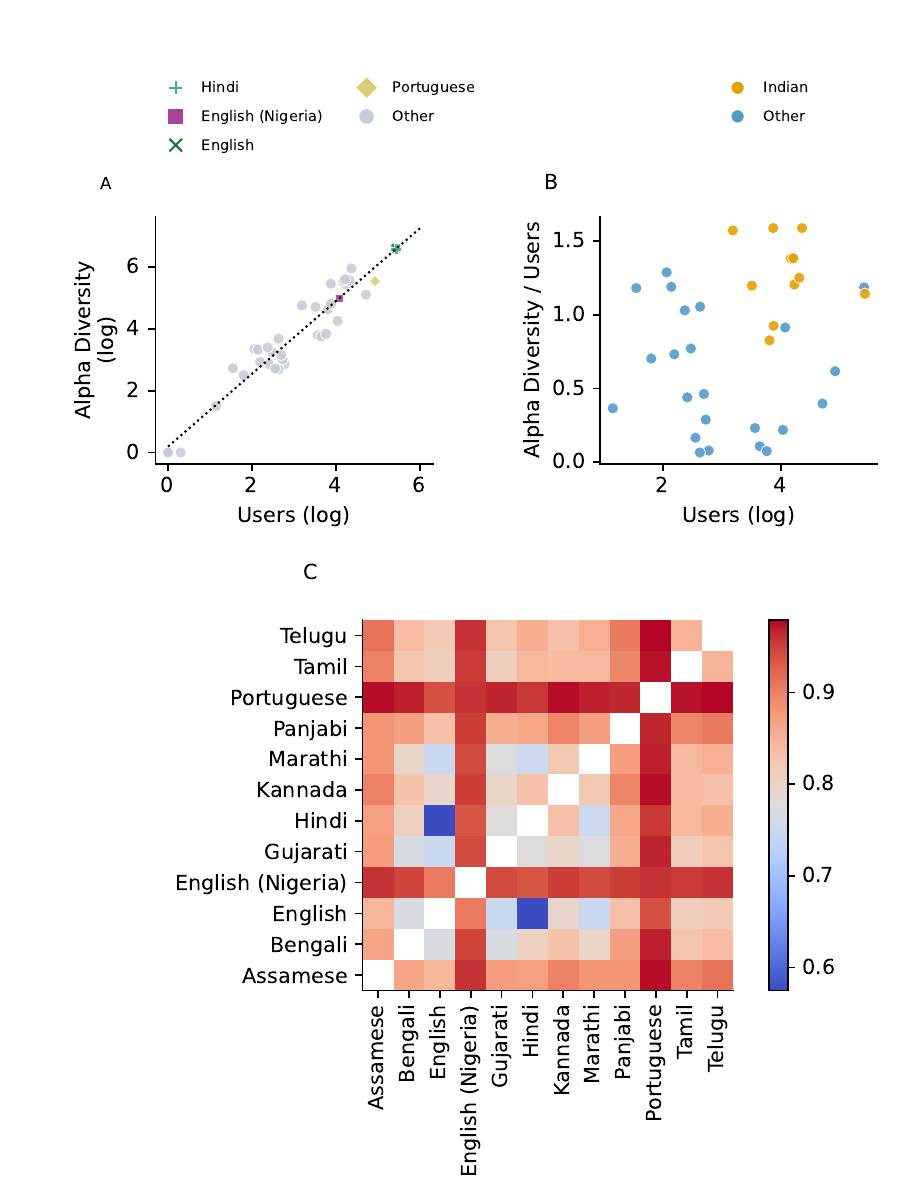}
    \caption{\textbf{Discourse richness and similarity across linguistic communities on Koo.} A) The alpha diversity of the discourse in a linguistic community, measured with the improved Chao1 estimator, plotted against the population size of the community, along with the linear fit (Spearman's $R^2 = 0.92$). Colours are used to indicate the main linguistic communities on Koo. B) The ratio between the improved Chao1 estimator and the population size plotted against the population size of the community. Colours indicate communities speaking an Indian language. C) The beta dissimilarity, measured with the Bray-Curtis index by considering the list of hashtags used by the largest linguistic communities on Koo and measuring their respective dissimilarity. Two communities with an index close to 0 use similar hashtags, whereas an index of 1 indicates that there is no overlap in the hashtags used by the communities.}
    \label{fig:alpha_beta_indian}
\end{figure}

Figure \ref{fig:alpha_beta_indian}A shows the improved Chao1 estimator for each linguistic community plotted against its population size. We measure a strong correlation ($R^2 = 0.92$) between the two variables, with bigger communities displaying a richer use of hashtags than smaller ones. There is, however, a non-negligible fluctuation in the measured richness for linguistic communities that share a similar population size, especially when looking at mid-sized communities.

To look more closely at this disparity, we display on figure \ref{fig:alpha_beta_indian}B the improved Chao1 estimator per user, plotted against the population size. Points are coloured to highlight the Indian official languages. Interestingly, we notice that for a similar population size, the ratio is higher for Indian languages than for non-Indian languages (with English, a language widely used in Indian administration, being the exception). 

These findings suggest that, with similar population sizes, Indian language speakers will give rise to a richer discourse on Koo than a non-Indian based linguistic community. These findings are especially insightful when considering that the Portuguese and Nigerian English speaking users are involved in far less rich conversations than their Indian peers, showcasing Koo's struggle to become a significant microblogging platform outside its native India \cite{dosunmu2023_nigeria}. 

The presence of a rich discourse among Indian communities does not necessarily indicate the existence of a national cohesion. The ``divide and rule'' policy, implemented by British imperials in India from 1857 to the country's independence in 1947, to sow divisions between ethnic groups in the country, led to long-standing cultural cleavages \cite{Seshia1998_divide} and to the emergence of a communal, rather than national, sense of identity \cite{bhagat2001_divide}. We can assess the influence of these divisions between communities on narrative content using the Bray-Curtis dissimilarity index \cite{Hao2019_bray}, a measure of beta diversity, defined as 
\begin{equation}
    \beta_{AB} = \frac{\sum_i |p_i^A - p_i^B|}{\sum_i (p_i^A - p_i^B)},
\end{equation}
where $p_i^A$ and $p_i^B$ represent the relative frequency of hashtag $i$ in the linguistic communities A and B, respectively. $\beta_{AB}$ provides us with a measure of the dissimilarity between the two environments: if the two linguistic communities use no similar hashtags, their dissimilarity $\beta_{AB}$ will be 1, whereas it will be equal to 0 if they use exactly the same hashtags.

Figure \ref{fig:alpha_beta_indian}C shows the Bray-Curtis dissimilarity between the discourse in two linguistic communities. We notice that the Portuguese and Nigerian communities have a high dissimilarity when compared to the other languages, which indicates that there is a low overlap in the hashtags shared by Portuguese and Nigerian English speakers with respect to other linguistic communities. This finding further confirms their status as isolated clusters. 

Looking at the Indian languages, we also notice a higher similarity shared between a few national languages, namely Gujarati, Bengali, Hindi and Marathi. These four languages are Indo-Aryan languages, thus sharing more linguistic similarities with each other than with Dravidian languages such as Telugu and Kannada. These results therefore outline the existence of a \textit{Sprachbund}, a set of languages that share many similarities in their structure \cite{emeneau1956_india}. This concept seeps within the online discourse on Koo, where we see a strong affinity between languages that share similar roots. Overall, Indian languages share a more similar discourse with each other than with non-Indian languages on the platform, which might explain why the conversation is richer amongst these communities.

\section*{Discussion \& Conclusion}

In this paper, we have analysed the emergence of a multi-lingual ecosystem on Koo, a microblogging platform based in India, which was shaped by successive migrations prompted by political and social events in India, Nigeria, and Brazil. This study follows our previous work on Koo where we discussed these migrations and released the Koo dataset \cite{mekacher2024_koo}. Here, we first looked at the impact of collective migrations to Koo in India, Nigeria and Brazil, and measured their respective success by comparing their impact on the daily registrations on Koo and their user retention over time. Second, we looked at the user interaction network and showed a strong linguistic segregation within isolated communities and the propensity for Hindi- and English-speaking users to be more central in the overall network. Third, we measured the average commitment and homophily for each linguistic community, and showed that linguistic communities with larger user-bases are more likely to be self-sufficient and dominantly display in-group interactions, with English being the exception due to its universal use across cultural communities. Fourth, we generated the global language network and highlighted the importance of linguistic crossovers between Indian languages. In contrast, Portuguese-speaking users only significantly overlap with English speaking users. Finally, we measured the richness and dissimilarity of linguistic clusters and noticed a strong interconnection across Indian languages, which also have a richer discourse than other communities with a comparable population size.

Our study offers a first insight into the emergence of a multi-lingual alt-tech platform, where the co-existence of diverse linguistic communities is driven by independent collective migrations. Despite Koo's ambitions to unite the non-English speaking world under a single banner, the language divide we measure both in terms of interactions and discourse similarity suggests that these communities are growing concurrently on the platform but without a strong tendency to overlap. While major languages can still thrive and be used by a sustainable community, less prevalent languages end up being marginally used, leading their speakers to be less involved in the major conversations taking place on the platform. These findings might indicate that minority linguistic communities are disenfranchised from the central political discourse on social media. Previous inquiries have already raised concerns about the access to many public services in India for local language speakers \cite{aula2014_language}. Koo currently supports 20 of India's 22 official languages, in a country with 122 languages listed in the national constitution.

Our analysis highlights the growth of the digital market in the Global South, which is often overlooked by the academic literature. Koo pledged to offer an online space catering to non-English speaking users and managed to attract key political figures from major emerging economies, leading the platform to become a serious competitor to Twitter \cite{chakravarty2023_koo}. Moreover, our research adds more nuance to the literature on alternative platforms that has been mostly focused on right-wing narratives. Koo is used by both the Indian and the Brazilian governments, representing contrasting political orientations. However, the Indian and Brazilian communities on Koo were shown to hardly interact with one another, indicating scarce engagements across the political spectrum. Koo's struggles to attract members of political parties other than the BJP and their allies in India highlights the difficulties alt-tech platforms face in building an online space for politically diverse communities \cite{cornish2022_koo}. This is further exacerbated by the recent rise of policies related to digital sovereignty across a number of countries, compromising Koo's ambition to become a unifying platform for communities outside the West \cite{larsen2022_sovereignty}. 

Our findings are limited by several factors, which can be explored in future studies. First, our analysis is restricted to Koo, which is the most popular microblogging platform based in India, and the second most popular India-based social platform in general (after ShareChat). Social platforms based in other nations in the Global South may display different structural patterns and should also be studies. Future studies may also consider extending our analysis to other popular platforms based in India including ShareChat, 2go and Line. Second, our analysis relies on the language identified by Koo's automated system for each post and comment on Koo. This may be problematic given that automatic language detection has various levels of reliability depending on the source language \cite{Hershcovich2022_nlp}. Further work may consider the accuracy of Koo's language identification pipeline, and whether it results in specific biases in our analysis. Finally, our work focuses on a static overview of Koo's interaction network by looking at all the user activity that took place over the four years studied. This ensures that we are able to fairly compare diverse linguistic communities who joined Koo at different times. However, an analysis of the temporal evolution of communities on Koo would be equally valuable, in particular to identify how the network evolved following each collective migration.

Overall, our study improves our understanding of social interaction patterns within multilingual communities by looking beyond western social platforms. We anticipate that our work will inspire more research on the global social media ecosystem, to ensure that our findings with respect to alternative platforms are nuanced by cultural and linguistic factors. Finally, our analysis stresses the need to develop a wider range of tools to analyse social media content in minority languages, many of which are currently underrepresented on the internet, and understudied by academics. 

\bibliography{apssamp}

\providecommand{\noopsort}[1]{}\providecommand{\singleletter}[1]{#1}%
\begin{thebibliography}{85}%
\makeatletter
\providecommand \@ifxundefined [1]{%
 \@ifx{#1\undefined}
}%
\providecommand \@ifnum [1]{%
 \ifnum #1\expandafter \@firstoftwo
 \else \expandafter \@secondoftwo
 \fi
}%
\providecommand \@ifx [1]{%
 \ifx #1\expandafter \@firstoftwo
 \else \expandafter \@secondoftwo
 \fi
}%
\providecommand \natexlab [1]{#1}%
\providecommand \enquote  [1]{``#1''}%
\providecommand \bibnamefont  [1]{#1}%
\providecommand \bibfnamefont [1]{#1}%
\providecommand \citenamefont [1]{#1}%
\providecommand \href@noop [0]{\@secondoftwo}%
\providecommand \href [0]{\begingroup \@sanitize@url \@href}%
\providecommand \@href[1]{\@@startlink{#1}\@@href}%
\providecommand \@@href[1]{\endgroup#1\@@endlink}%
\providecommand \@sanitize@url [0]{\catcode `\\12\catcode `\$12\catcode `\&12\catcode `\#12\catcode `\^12\catcode `\_12\catcode `\%12\relax}%
\providecommand \@@startlink[1]{}%
\providecommand \@@endlink[0]{}%
\providecommand \url  [0]{\begingroup\@sanitize@url \@url }%
\providecommand \@url [1]{\endgroup\@href {#1}{\urlprefix }}%
\providecommand \urlprefix  [0]{URL }%
\providecommand \Eprint [0]{\href }%
\providecommand \doibase [0]{http://dx.doi.org/}%
\providecommand \selectlanguage [0]{\@gobble}%
\providecommand \bibinfo  [0]{\@secondoftwo}%
\providecommand \bibfield  [0]{\@secondoftwo}%
\providecommand \translation [1]{[#1]}%
\providecommand \BibitemOpen [0]{}%
\providecommand \bibitemStop [0]{}%
\providecommand \bibitemNoStop [0]{.\EOS\space}%
\providecommand \EOS [0]{\spacefactor3000\relax}%
\providecommand \BibitemShut  [1]{\csname bibitem#1\endcsname}%
\let\auto@bib@innerbib\@empty
\bibitem [{\citenamefont {Wike}\ \emph {et~al.}(2022)\citenamefont {Wike}, \citenamefont {Silver}, \citenamefont {Fetterolf}, \citenamefont {Huang}, \citenamefont {Austin}, \citenamefont {Clancy},\ and\ \citenamefont {Gubbala}}]{wike2022_democracy}%
  \BibitemOpen
  \bibfield  {author} {\bibinfo {author} {\bibfnamefont {R.}~\bibnamefont {Wike}}, \bibinfo {author} {\bibfnamefont {L.}~\bibnamefont {Silver}}, \bibinfo {author} {\bibfnamefont {J.}~\bibnamefont {Fetterolf}}, \bibinfo {author} {\bibfnamefont {C.}~\bibnamefont {Huang}}, \bibinfo {author} {\bibfnamefont {S.}~\bibnamefont {Austin}}, \bibinfo {author} {\bibfnamefont {L.}~\bibnamefont {Clancy}}, \ and\ \bibinfo {author} {\bibfnamefont {S.}~\bibnamefont {Gubbala}},\ }\href@noop {} {\enquote {\bibinfo {title} {Social media seen as mostly good for democracy across many nations, but u.s. is a major outlier},}\ }\bibinfo {howpublished} {\url{https://www.pewresearch.org/global/2022/12/06/social-media-seen-as-mostly-good-for-democracy-across-many-nations-but-u-s-is-a-major-outlier/} Accessed 8 February, 2024} (\bibinfo {year} {2022})\BibitemShut {NoStop}%
\bibitem [{\citenamefont {Poushter}\ \emph {et~al.}(2018)\citenamefont {Poushter}, \citenamefont {Bishop},\ and\ \citenamefont {Chwe}}]{poushter2018_social}%
  \BibitemOpen
  \bibfield  {author} {\bibinfo {author} {\bibfnamefont {J.}~\bibnamefont {Poushter}}, \bibinfo {author} {\bibfnamefont {C.}~\bibnamefont {Bishop}}, \ and\ \bibinfo {author} {\bibfnamefont {H.}~\bibnamefont {Chwe}},\ }\href@noop {} {\enquote {\bibinfo {title} {Social media use continues to rise in developing countries but plateaus across developed ones},}\ }\bibinfo {howpublished} {\url{https://www.pewresearch.org/global/2018/06/19/social-media-use-continues-to-rise-in-developing-countries-but-plateaus-across-developed-ones/} Accessed 2 February, 2024} (\bibinfo {year} {2018})\BibitemShut {NoStop}%
\bibitem [{\citenamefont {Mandavia}\ and\ \citenamefont {Peermohamed}(2021)}]{mandavia2021_vernacular}%
  \BibitemOpen
  \bibfield  {author} {\bibinfo {author} {\bibfnamefont {M.}~\bibnamefont {Mandavia}}\ and\ \bibinfo {author} {\bibfnamefont {A.}~\bibnamefont {Peermohamed}},\ }\href@noop {} {\enquote {\bibinfo {title} {Facebook disputes claims of inadequate flagging of vernacular content},}\ }\bibinfo {howpublished} {\url{https://economictimes.indiatimes.com/tech/technology/facebook-disputes-claims-of-inadequate-flagging-of-vernacular-content/articleshow/86814892.cms?from=mdr} Accessed 2 February, 2024} (\bibinfo {year} {2021})\BibitemShut {NoStop}%
\bibitem [{\citenamefont {Leong}(2022)}]{leong2022_asia}%
  \BibitemOpen
  \bibfield  {author} {\bibinfo {author} {\bibfnamefont {P.}~\bibnamefont {Leong}},\ }\href@noop {} {\enquote {\bibinfo {title} {Tech giants grapple with content moderation in south-east asia},}\ }\bibinfo {howpublished} {\url{https://www.straitstimes.com/opinion/tech-giants-grapple-with-content-moderation-in-south-east-asia} Accessed 7 February, 2024} (\bibinfo {year} {2022})\BibitemShut {NoStop}%
\bibitem [{\citenamefont {Milmo}(2021)}]{milmo2021_facebook}%
  \BibitemOpen
  \bibfield  {author} {\bibinfo {author} {\bibfnamefont {D.}~\bibnamefont {Milmo}},\ }\href@noop {} {\enquote {\bibinfo {title} {Facebook revelations: what is in cache of internal documents?}}\ }\bibinfo {howpublished} {\url{https://www.theguardian.com/technology/2021/oct/25/facebook-revelations-from-misinformation-to-mental-health} Accessed 12 February, 2024} (\bibinfo {year} {2021})\BibitemShut {NoStop}%
\bibitem [{\citenamefont {Statista}(2023)}]{statista2023_instagram}%
  \BibitemOpen
  \bibfield  {author} {\bibinfo {author} {\bibnamefont {Statista}},\ }\href@noop {} {\enquote {\bibinfo {title} {Leading countries based on instagram audience size as of january 2023},}\ }\bibinfo {howpublished} {\url{https://www.statista.com/statistics/578364/countries-with-most-instagram-users/} Accessed 18 August, 2023} (\bibinfo {year} {2023})\BibitemShut {NoStop}%
\bibitem [{\citenamefont {Worldwide}(2022)}]{mbw2022_youtube}%
  \BibitemOpen
  \bibfield  {author} {\bibinfo {author} {\bibfnamefont {M.~B.}\ \bibnamefont {Worldwide}},\ }\href@noop {} {\enquote {\bibinfo {title} {India is home to the largest number of youtube viewers in the world. but the platform’s music chart isn’t without flaws},}\ }\bibinfo {howpublished} {\url{https://www.musicbusinessworldwide.com/india-is-home-to-the-largest-number-of-youtube-viewers-in-the-world-but-the-platforms/} Accessed 31 October, 2023} (\bibinfo {year} {2022})\BibitemShut {NoStop}%
\bibitem [{\citenamefont {Vengattil}\ and\ \citenamefont {Kalra}(2022)}]{vengattil2022_facebook}%
  \BibitemOpen
  \bibfield  {author} {\bibinfo {author} {\bibfnamefont {M.}~\bibnamefont {Vengattil}}\ and\ \bibinfo {author} {\bibfnamefont {A.}~\bibnamefont {Kalra}},\ }\href@noop {} {\enquote {\bibinfo {title} {Facebook's growth woes in india: too much nudity, not enough women},}\ }\bibinfo {howpublished} {\url{https://www.reuters.com/article/meta-india-facebook-insight-idCAKBN2OW0FG} Accessed 31 October, 2023} (\bibinfo {year} {2022})\BibitemShut {NoStop}%
\bibitem [{\citenamefont {Radhakrishna}(2022)}]{radhakrishna2022_koo}%
  \BibitemOpen
  \bibfield  {author} {\bibinfo {author} {\bibfnamefont {A.}~\bibnamefont {Radhakrishna}},\ }\href@noop {} {\enquote {\bibinfo {title} {Multilingual social media could catalyse digital inclusion in india},}\ }\bibinfo {howpublished} {\url{https://www.livemint.com/opinion/online-views/multilingual-social-media-could-catalyse-digital-inclusion-in-india-11656950158510.html} Accessed 22 January, 2024} (\bibinfo {year} {2022})\BibitemShut {NoStop}%
\bibitem [{\citenamefont {Platform}(2024)}]{twitter2024_languages}%
  \BibitemOpen
  \bibfield  {author} {\bibinfo {author} {\bibfnamefont {X.~D.}\ \bibnamefont {Platform}},\ }\href@noop {} {\enquote {\bibinfo {title} {Supported languages and browsers},}\ }\bibinfo {howpublished} {\url{https://developer.twitter.com/en/docs/twitter-for-websites/supported-languages} Accessed 11 March, 2024} (\bibinfo {year} {2024})\BibitemShut {NoStop}%
\bibitem [{\citenamefont {Singh}(2022)}]{singh2022_koo}%
  \BibitemOpen
  \bibfield  {author} {\bibinfo {author} {\bibfnamefont {S.}~\bibnamefont {Singh}},\ }\href@noop {} {\enquote {\bibinfo {title} {Koo said to be the second largest microblogging platform in the world now},}\ }\bibinfo {howpublished} {\url{https://www.cnbctv18.com/business/companies/koo-becomes-second-largest-microblogging-platform-in-the-world-after-twitter-15179881.htm} Accessed 8 December, 2023} (\bibinfo {year} {2022})\BibitemShut {NoStop}%
\bibitem [{\citenamefont {Inamdar}(2022)}]{inamdar2022_koo}%
  \BibitemOpen
  \bibfield  {author} {\bibinfo {author} {\bibfnamefont {N.}~\bibnamefont {Inamdar}},\ }\href@noop {} {\enquote {\bibinfo {title} {Koo: India's twitter alternative with global ambitions},}\ }\bibinfo {howpublished} {\url{https://www.bbc.co.uk/news/world-asia-india-60194920} Accessed 14 September, 2023} (\bibinfo {year} {2022})\BibitemShut {NoStop}%
\bibitem [{\citenamefont {Mekacher}(2024)}]{mekacher2024_zenodo}%
  \BibitemOpen
  \bibfield  {author} {\bibinfo {author} {\bibfnamefont {A.}~\bibnamefont {Mekacher}},\ }\href {\doibase 10.5281/zenodo.10476212} {\enquote {\bibinfo {title} {{The Koo Dataset: An Indian Microblogging Platform With Global Ambitions}},}\ } (\bibinfo {year} {2024})\BibitemShut {NoStop}%
\bibitem [{\citenamefont {Mekacher}\ \emph {et~al.}(2024)\citenamefont {Mekacher}, \citenamefont {Falkenberg},\ and\ \citenamefont {Baronchelli}}]{mekacher2024_koo}%
  \BibitemOpen
  \bibfield  {author} {\bibinfo {author} {\bibfnamefont {A.}~\bibnamefont {Mekacher}}, \bibinfo {author} {\bibfnamefont {M.}~\bibnamefont {Falkenberg}}, \ and\ \bibinfo {author} {\bibfnamefont {A.}~\bibnamefont {Baronchelli}},\ }\href {\doibase 10.48550/ARXIV.2401.07599} {\enquote {\bibinfo {title} {The koo dataset: An indian microblogging platform with global ambitions},}\ } (\bibinfo {year} {2024})\BibitemShut {NoStop}%
\bibitem [{\citenamefont {Mekacher}\ and\ \citenamefont {Papasavva}(2022)}]{mekacher_2022voat}%
  \BibitemOpen
  \bibfield  {author} {\bibinfo {author} {\bibfnamefont {A.}~\bibnamefont {Mekacher}}\ and\ \bibinfo {author} {\bibfnamefont {A.}~\bibnamefont {Papasavva}},\ }\href {https://ojs.aaai.org/index.php/ICWSM/article/view/19382} {\bibfield  {journal} {\bibinfo  {journal} {Proceedings of the International AAAI Conference on Web and Social Media}\ }\textbf {\bibinfo {volume} {16}},\ \bibinfo {pages} {1302} (\bibinfo {year} {2022})}\BibitemShut {NoStop}%
\bibitem [{\citenamefont {Ali}\ \emph {et~al.}(2021)\citenamefont {Ali}, \citenamefont {Saeed}, \citenamefont {Aldreabi}, \citenamefont {Blackburn}, \citenamefont {De~Cristofaro}, \citenamefont {Zannettou},\ and\ \citenamefont {Stringhini}}]{ali2021_deplatforming}%
  \BibitemOpen
  \bibfield  {author} {\bibinfo {author} {\bibfnamefont {S.}~\bibnamefont {Ali}}, \bibinfo {author} {\bibfnamefont {M.~H.}\ \bibnamefont {Saeed}}, \bibinfo {author} {\bibfnamefont {E.}~\bibnamefont {Aldreabi}}, \bibinfo {author} {\bibfnamefont {J.}~\bibnamefont {Blackburn}}, \bibinfo {author} {\bibfnamefont {E.}~\bibnamefont {De~Cristofaro}}, \bibinfo {author} {\bibfnamefont {S.}~\bibnamefont {Zannettou}}, \ and\ \bibinfo {author} {\bibfnamefont {G.}~\bibnamefont {Stringhini}},\ }in\ \href {\doibase 10.1145/3447535.3462637} {\emph {\bibinfo {booktitle} {13th ACM Web Science Conference 2021}}}\ (\bibinfo  {publisher} {Association for Computing Machinery},\ \bibinfo {address} {New York, NY, USA},\ \bibinfo {year} {2021})\BibitemShut {NoStop}%
\bibitem [{\citenamefont {Papasavva}\ and\ \citenamefont {Mariconti}(2023)}]{papasavva2023_poal}%
  \BibitemOpen
  \bibfield  {author} {\bibinfo {author} {\bibfnamefont {A.}~\bibnamefont {Papasavva}}\ and\ \bibinfo {author} {\bibfnamefont {E.}~\bibnamefont {Mariconti}},\ }\href {\doibase 10.48550/ARXIV.2302.01397} {\enquote {\bibinfo {title} {Waiting for q: An exploration of qanon users' online migration to poal in the wake of voat's demise},}\ } (\bibinfo {year} {2023})\BibitemShut {NoStop}%
\bibitem [{\citenamefont {Hong}\ \emph {et~al.}(2021)\citenamefont {Hong}, \citenamefont {Convertino},\ and\ \citenamefont {Chi}}]{Hong2021_lang}%
  \BibitemOpen
  \bibfield  {author} {\bibinfo {author} {\bibfnamefont {L.}~\bibnamefont {Hong}}, \bibinfo {author} {\bibfnamefont {G.}~\bibnamefont {Convertino}}, \ and\ \bibinfo {author} {\bibfnamefont {E.}~\bibnamefont {Chi}},\ }\href {\doibase 10.1609/icwsm.v5i1.14184} {\bibfield  {journal} {\bibinfo  {journal} {Proceedings of the International {AAAI} Conference on Web and Social Media}\ }\textbf {\bibinfo {volume} {5}},\ \bibinfo {pages} {518} (\bibinfo {year} {2021})}\BibitemShut {NoStop}%
\bibitem [{\citenamefont {Mocanu}\ \emph {et~al.}(2013)\citenamefont {Mocanu}, \citenamefont {Baronchelli}, \citenamefont {Perra}, \citenamefont {Gon{\c{c}}alves}, \citenamefont {Zhang},\ and\ \citenamefont {Vespignani}}]{Mocanu2013_lang}%
  \BibitemOpen
  \bibfield  {author} {\bibinfo {author} {\bibfnamefont {D.}~\bibnamefont {Mocanu}}, \bibinfo {author} {\bibfnamefont {A.}~\bibnamefont {Baronchelli}}, \bibinfo {author} {\bibfnamefont {N.}~\bibnamefont {Perra}}, \bibinfo {author} {\bibfnamefont {B.}~\bibnamefont {Gon{\c{c}}alves}}, \bibinfo {author} {\bibfnamefont {Q.}~\bibnamefont {Zhang}}, \ and\ \bibinfo {author} {\bibfnamefont {A.}~\bibnamefont {Vespignani}},\ }\href {\doibase 10.1371/journal.pone.0061981} {\bibfield  {journal} {\bibinfo  {journal} {{PLoS} {ONE}}\ }\textbf {\bibinfo {volume} {8}},\ \bibinfo {pages} {e61981} (\bibinfo {year} {2013})}\BibitemShut {NoStop}%
\bibitem [{\citenamefont {Krishnamurthy}\ \emph {et~al.}(2008)\citenamefont {Krishnamurthy}, \citenamefont {Gill},\ and\ \citenamefont {Arlitt}}]{Krishnamurthy2008_twitter}%
  \BibitemOpen
  \bibfield  {author} {\bibinfo {author} {\bibfnamefont {B.}~\bibnamefont {Krishnamurthy}}, \bibinfo {author} {\bibfnamefont {P.}~\bibnamefont {Gill}}, \ and\ \bibinfo {author} {\bibfnamefont {M.}~\bibnamefont {Arlitt}},\ }in\ \href {\doibase 10.1145/1397735.1397741} {\emph {\bibinfo {booktitle} {Proceedings of the first workshop on Online social networks}}}\ (\bibinfo  {publisher} {{ACM}},\ \bibinfo {year} {2008})\BibitemShut {NoStop}%
\bibitem [{\citenamefont {Java}\ \emph {et~al.}(2007)\citenamefont {Java}, \citenamefont {Song}, \citenamefont {Finin},\ and\ \citenamefont {Tseng}}]{Java2007_twitter}%
  \BibitemOpen
  \bibfield  {author} {\bibinfo {author} {\bibfnamefont {A.}~\bibnamefont {Java}}, \bibinfo {author} {\bibfnamefont {X.}~\bibnamefont {Song}}, \bibinfo {author} {\bibfnamefont {T.}~\bibnamefont {Finin}}, \ and\ \bibinfo {author} {\bibfnamefont {B.}~\bibnamefont {Tseng}},\ }in\ \href {\doibase 10.1145/1348549.1348556} {\emph {\bibinfo {booktitle} {Proceedings of the 9th {WebKDD} and 1st {SNA}-{KDD} 2007 workshop on Web mining and social network analysis}}}\ (\bibinfo  {publisher} {{ACM}},\ \bibinfo {year} {2007})\BibitemShut {NoStop}%
\bibitem [{\citenamefont {Weerkamp}\ \emph {et~al.}(2011)\citenamefont {Weerkamp}, \citenamefont {Carter},\ and\ \citenamefont {Tsagkias}}]{weerkamp2011_lang}%
  \BibitemOpen
  \bibfield  {author} {\bibinfo {author} {\bibfnamefont {W.}~\bibnamefont {Weerkamp}}, \bibinfo {author} {\bibfnamefont {S.}~\bibnamefont {Carter}}, \ and\ \bibinfo {author} {\bibfnamefont {M.}~\bibnamefont {Tsagkias}},\ }\href@noop {} {\bibfield  {journal} {\bibinfo  {journal} {Interdisciplinary Journal for the Study of Discourse}\ } (\bibinfo {year} {2011})}\BibitemShut {NoStop}%
\bibitem [{\citenamefont {Takhteyev}\ \emph {et~al.}(2012)\citenamefont {Takhteyev}, \citenamefont {Gruzd},\ and\ \citenamefont {Wellman}}]{Takhteyev2012_language}%
  \BibitemOpen
  \bibfield  {author} {\bibinfo {author} {\bibfnamefont {Y.}~\bibnamefont {Takhteyev}}, \bibinfo {author} {\bibfnamefont {A.}~\bibnamefont {Gruzd}}, \ and\ \bibinfo {author} {\bibfnamefont {B.}~\bibnamefont {Wellman}},\ }\href {\doibase 10.1016/j.socnet.2011.05.006} {\bibfield  {journal} {\bibinfo  {journal} {Social Networks}\ }\textbf {\bibinfo {volume} {34}},\ \bibinfo {pages} {73} (\bibinfo {year} {2012})}\BibitemShut {NoStop}%
\bibitem [{\citenamefont {Falkenberg}\ \emph {et~al.}(2023)\citenamefont {Falkenberg}, \citenamefont {Zollo}, \citenamefont {Quattrociocchi}, \citenamefont {Pfeffer},\ and\ \citenamefont {Baronchelli}}]{falkenberg2023_polarization}%
  \BibitemOpen
  \bibfield  {author} {\bibinfo {author} {\bibfnamefont {M.}~\bibnamefont {Falkenberg}}, \bibinfo {author} {\bibfnamefont {F.}~\bibnamefont {Zollo}}, \bibinfo {author} {\bibfnamefont {W.}~\bibnamefont {Quattrociocchi}}, \bibinfo {author} {\bibfnamefont {J.}~\bibnamefont {Pfeffer}}, \ and\ \bibinfo {author} {\bibfnamefont {A.}~\bibnamefont {Baronchelli}},\ }\href {\doibase 10.48550/ARXIV.2311.18535} {\bibfield  {journal} {\bibinfo  {journal} {ArXiv Preprint}\ ,\ \bibinfo {pages} {2311.18535}} (\bibinfo {year} {2023})}\BibitemShut {NoStop}%
\bibitem [{\citenamefont {Kim}\ \emph {et~al.}(2014)\citenamefont {Kim}, \citenamefont {Weber}, \citenamefont {Wei},\ and\ \citenamefont {Oh}}]{Kim2014_multilingual}%
  \BibitemOpen
  \bibfield  {author} {\bibinfo {author} {\bibfnamefont {S.}~\bibnamefont {Kim}}, \bibinfo {author} {\bibfnamefont {I.}~\bibnamefont {Weber}}, \bibinfo {author} {\bibfnamefont {L.}~\bibnamefont {Wei}}, \ and\ \bibinfo {author} {\bibfnamefont {A.}~\bibnamefont {Oh}},\ }in\ \href {\doibase 10.1145/2631775.2631824} {\emph {\bibinfo {booktitle} {Proceedings of the 25th ACM conference on Hypertext and social media}}},\ \bibinfo {series and number} {HT ’14}\ (\bibinfo  {publisher} {ACM},\ \bibinfo {year} {2014})\BibitemShut {NoStop}%
\bibitem [{\citenamefont {Androutsopoulos}(2007)}]{Androutsopoulos2007_language}%
  \BibitemOpen
  \bibfield  {author} {\bibinfo {author} {\bibfnamefont {J.}~\bibnamefont {Androutsopoulos}},\ }\enquote {\bibinfo {title} {Language choice and code switching in german-based diasporic web forums},}\ in\ \href {\doibase 10.1093/acprof:oso/9780195304794.003.0015} {\emph {\bibinfo {booktitle} {The Multilingual Internet}}}\ (\bibinfo  {publisher} {Oxford University Press},\ \bibinfo {year} {2007})\ p.\ \bibinfo {pages} {340–361}\BibitemShut {NoStop}%
\bibitem [{\citenamefont {Bailey}\ \emph {et~al.}(2013)\citenamefont {Bailey}, \citenamefont {Goggins},\ and\ \citenamefont {Ingham}}]{bailey2013_language}%
  \BibitemOpen
  \bibfield  {author} {\bibinfo {author} {\bibfnamefont {G.}~\bibnamefont {Bailey}}, \bibinfo {author} {\bibfnamefont {J.}~\bibnamefont {Goggins}}, \ and\ \bibinfo {author} {\bibfnamefont {T.}~\bibnamefont {Ingham}},\ }\href@noop {} {\bibfield  {journal} {\bibinfo  {journal} {Report by Multilingual Manchester. School of Languages, Linguistics and Cultures at the University of Manchester. http://bit. ly/2kG42Qf}\ } (\bibinfo {year} {2013})}\BibitemShut {NoStop}%
\bibitem [{\citenamefont {Lin}\ \emph {et~al.}(2023)\citenamefont {Lin}, \citenamefont {Wu},\ and\ \citenamefont {Mason}}]{Lin2023_facebook}%
  \BibitemOpen
  \bibfield  {author} {\bibinfo {author} {\bibfnamefont {Y.-R.}\ \bibnamefont {Lin}}, \bibinfo {author} {\bibfnamefont {S.}~\bibnamefont {Wu}}, \ and\ \bibinfo {author} {\bibfnamefont {W.}~\bibnamefont {Mason}},\ }\href {\doibase 10.1140/epjds/s13688-023-00388-4} {\bibfield  {journal} {\bibinfo  {journal} {{EPJ} Data Science}\ }\textbf {\bibinfo {volume} {12}} (\bibinfo {year} {2023}),\ 10.1140/epjds/s13688-023-00388-4}\BibitemShut {NoStop}%
\bibitem [{\citenamefont {Apte}(1976)}]{apte1976_multilingualism}%
  \BibitemOpen
  \bibfield  {author} {\bibinfo {author} {\bibfnamefont {M.~L.}\ \bibnamefont {Apte}},\ }\href@noop {} {\bibfield  {journal} {\bibinfo  {journal} {Language and Politics}\ ,\ \bibinfo {pages} {141}} (\bibinfo {year} {1976})}\BibitemShut {NoStop}%
\bibitem [{\citenamefont {Mallikarjun}(2010)}]{mallikarjun2010_patterns}%
  \BibitemOpen
  \bibfield  {author} {\bibinfo {author} {\bibfnamefont {B.}~\bibnamefont {Mallikarjun}},\ }\href@noop {} {\bibfield  {journal} {\bibinfo  {journal} {Language in India}\ }\textbf {\bibinfo {volume} {10}},\ \bibinfo {pages} {1} (\bibinfo {year} {2010})}\BibitemShut {NoStop}%
\bibitem [{\citenamefont {Pattanayak}(1990)}]{pattanayak1990_multilingualism}%
  \BibitemOpen
  \bibfield  {author} {\bibinfo {author} {\bibfnamefont {D.~P.}\ \bibnamefont {Pattanayak}},\ }\href@noop {} {\emph {\bibinfo {title} {Multilingualism in India}}},\ \bibinfo {number} {61}\ (\bibinfo  {publisher} {Multilingual Matters},\ \bibinfo {year} {1990})\BibitemShut {NoStop}%
\bibitem [{\citenamefont {Kari}(2002)}]{kari2002_multilingualism}%
  \BibitemOpen
  \bibfield  {author} {\bibinfo {author} {\bibfnamefont {E.~E.}\ \bibnamefont {Kari}},\ }in\ \href@noop {} {\emph {\bibinfo {booktitle} {A Paper Presented at the Seminar on Multilingual Situation and Related Local Cultures in Asia and Africa Institute for the Study of Languages and Cultures of Asia and Africa, Tokyo University of Foreign Studies, Tokyo}}},\ Vol.~\bibinfo {volume} {25}\ (\bibinfo {year} {2002})\BibitemShut {NoStop}%
\bibitem [{\citenamefont {Moses~Omoniyi}(2012)}]{moses2012_nigeria}%
  \BibitemOpen
  \bibfield  {author} {\bibinfo {author} {\bibfnamefont {A.}~\bibnamefont {Moses~Omoniyi}},\ }\href {\doibase 10.5539/elt.v5n10p12} {\bibfield  {journal} {\bibinfo  {journal} {English Language Teaching}\ }\textbf {\bibinfo {volume} {5}} (\bibinfo {year} {2012}),\ 10.5539/elt.v5n10p12}\BibitemShut {NoStop}%
\bibitem [{\citenamefont {Ellis-Petersen}(2023)}]{ellis2023_twitter}%
  \BibitemOpen
  \bibfield  {author} {\bibinfo {author} {\bibfnamefont {H.}~\bibnamefont {Ellis-Petersen}},\ }\href@noop {} {\enquote {\bibinfo {title} {India threatened to shut twitter down, co-founder jack dorsey says},}\ }\bibinfo {howpublished} {\url{https://www.theguardian.com/world/2023/jun/13/india-threatened-to-shut-twitter-down-co-founder-jack-dorsey-says} Accessed 20 October, 2023} (\bibinfo {year} {2023})\BibitemShut {NoStop}%
\bibitem [{\citenamefont {Shih}(2021)}]{gerry2021_farmer}%
  \BibitemOpen
  \bibfield  {author} {\bibinfo {author} {\bibfnamefont {G.}~\bibnamefont {Shih}},\ }\href@noop {} {\enquote {\bibinfo {title} {In india, a government-friendly social media network challenges twitter},}\ }\bibinfo {howpublished} {\url{https://www.washingtonpost.com/world/2021/11/16/india-twitter-koo-social-network/} Accessed 15 August, 2023} (\bibinfo {year} {2021})\BibitemShut {NoStop}%
\bibitem [{\citenamefont {Nilesh}(2021)}]{nilesh2021_koo}%
  \BibitemOpen
  \bibfield  {author} {\bibinfo {author} {\bibfnamefont {C.}~\bibnamefont {Nilesh}},\ }\href@noop {} {\enquote {\bibinfo {title} {How koo became india’s hindu nationalist–approved twitter alternative},}\ }\bibinfo {howpublished} {\url{https://restofworld.org/2021/how-koo-became-a-right-wing-darling-in-india/} Accessed 15 August, 2023} (\bibinfo {year} {2021})\BibitemShut {NoStop}%
\bibitem [{\citenamefont {Ayomide}(2022)}]{ayomide2022_nigeria}%
  \BibitemOpen
  \bibfield  {author} {\bibinfo {author} {\bibfnamefont {A.}~\bibnamefont {Ayomide}},\ }\href@noop {} {\enquote {\bibinfo {title} {Inside indian app koo, where nigerians are migrating to after a twitter ban},}\ }\bibinfo {howpublished} {\url{https://qz.com/africa/2024107/nigerian-government-moves-to-indian-app-koo-after-twitter-ban} Accessed 15 August, 2023} (\bibinfo {year} {2022})\BibitemShut {NoStop}%
\bibitem [{\citenamefont {Princewill}\ and\ \citenamefont {Busari}(2021)}]{princewill2021_nigeria}%
  \BibitemOpen
  \bibfield  {author} {\bibinfo {author} {\bibfnamefont {N.}~\bibnamefont {Princewill}}\ and\ \bibinfo {author} {\bibfnamefont {S.}~\bibnamefont {Busari}},\ }\href@noop {} {\enquote {\bibinfo {title} {Nigeria bans twitter after company deletes president buhari’s tweet},}\ }\bibinfo {howpublished} {\url{https://edition.cnn.com/2021/06/04/africa/nigeria-suspends-twitter-operations-intl/index.html/} Accessed 31 October, 2023} (\bibinfo {year} {2021})\BibitemShut {NoStop}%
\bibitem [{\citenamefont {Abubakar}\ and\ \citenamefont {Nilesh}(2021)}]{abubakar2021_nigeria}%
  \BibitemOpen
  \bibfield  {author} {\bibinfo {author} {\bibfnamefont {I.}~\bibnamefont {Abubakar}}\ and\ \bibinfo {author} {\bibfnamefont {C.}~\bibnamefont {Nilesh}},\ }\href@noop {} {\enquote {\bibinfo {title} {Koo is selling itself as a twitter substitute in nigeria},}\ }\bibinfo {howpublished} {\url{https://restofworld.org/2021/koo-is-selling-itself-as-a-twitter-substitute-in-nigeria/} Accessed 15 August, 2023} (\bibinfo {year} {2021})\BibitemShut {NoStop}%
\bibitem [{\citenamefont {Phartiyal}\ and\ \citenamefont {Kalra}(2021)}]{phartiyal2021_india}%
  \BibitemOpen
  \bibfield  {author} {\bibinfo {author} {\bibfnamefont {S.}~\bibnamefont {Phartiyal}}\ and\ \bibinfo {author} {\bibfnamefont {A.}~\bibnamefont {Kalra}},\ }\href@noop {} {\enquote {\bibinfo {title} {In spats with twitter, india's government begins messaging shift to rival koo},}\ }\bibinfo {howpublished} {\url{https://www.reuters.com/world/india/spats-with-twitter-indias-government-begins-messaging-shift-rival-koo-2021-07-28/} Accessed 15 August, 2023} (\bibinfo {year} {2021})\BibitemShut {NoStop}%
\bibitem [{\citenamefont {Mekacher}\ \emph {et~al.}(2023)\citenamefont {Mekacher}, \citenamefont {Falkenberg},\ and\ \citenamefont {Baronchelli}}]{mekacher2023_gettr}%
  \BibitemOpen
  \bibfield  {author} {\bibinfo {author} {\bibfnamefont {A.}~\bibnamefont {Mekacher}}, \bibinfo {author} {\bibfnamefont {M.}~\bibnamefont {Falkenberg}}, \ and\ \bibinfo {author} {\bibfnamefont {A.}~\bibnamefont {Baronchelli}},\ }\href {\doibase 10.1093/pnasnexus/pgad346} {\bibfield  {journal} {\bibinfo  {journal} {PNAS Nexus}\ }\textbf {\bibinfo {volume} {2}} (\bibinfo {year} {2023}),\ 10.1093/pnasnexus/pgad346}\BibitemShut {NoStop}%
\bibitem [{\citenamefont {González~Ormerod}(2022)}]{gonzalez2022_brazil}%
  \BibitemOpen
  \bibfield  {author} {\bibinfo {author} {\bibfnamefont {A.}~\bibnamefont {González~Ormerod}},\ }\href@noop {} {\enquote {\bibinfo {title} {Twitter drama has brazilians flocking to indian platform koo},}\ }\bibinfo {howpublished} {\url{https://restofworld.org/2022/twitter-brazil-koo/} Accessed 16 August, 2023} (\bibinfo {year} {2022})\BibitemShut {NoStop}%
\bibitem [{\citenamefont {Mihindukulasuriya}(2022)}]{regina2022_lula}%
  \BibitemOpen
  \bibfield  {author} {\bibinfo {author} {\bibfnamefont {R.}~\bibnamefont {Mihindukulasuriya}},\ }\href@noop {} {\enquote {\bibinfo {title} {Brazilian president lula joins india’s twitter rival koo, gets over 50,000 followers in 4 hours},}\ }\bibinfo {howpublished} {\url{https://theprint.in/world/brazilian-president-lula-joins-indias-twitter-rival-koo-gets-over-50000-followers-in-4-hours/1245516/} Accessed 23 October, 2023} (\bibinfo {year} {2022})\BibitemShut {NoStop}%
\bibitem [{\citenamefont {Reuben~Das}(2022)}]{reuben2022_brazil}%
  \BibitemOpen
  \bibfield  {author} {\bibinfo {author} {\bibfnamefont {M.}~\bibnamefont {Reuben~Das}},\ }\href@noop {} {\enquote {\bibinfo {title} {India’s koo launched in brazil in portuguese, becomes the top downloaded app in 48 hours},}\ }\bibinfo {howpublished} {\url{https://www.firstpost.com/tech/news-analysis/koo-launched-in-brazil-in-portuguese-becomes-the-top-downloaded-app-in-48-hours-11671621.html} Accessed 24 January, 2024} (\bibinfo {year} {2022})\BibitemShut {NoStop}%
\bibitem [{\citenamefont {Akinwotu}(2022)}]{akinwotu2022_nigeria}%
  \BibitemOpen
  \bibfield  {author} {\bibinfo {author} {\bibfnamefont {E.}~\bibnamefont {Akinwotu}},\ }\href@noop {} {\enquote {\bibinfo {title} {Nigeria lifts twitter ban seven months after site deleted president’s post},}\ }\bibinfo {howpublished} {\url{https://www.theguardian.com/world/2022/jan/13/nigeria-lifts-twitter-ban-seven-months-after-site-deleted-presidents-post/} Accessed 2 November, 2023} (\bibinfo {year} {2022})\BibitemShut {NoStop}%
\bibitem [{\citenamefont {Dosunmu}(2023)}]{dosunmu2023_nigeria}%
  \BibitemOpen
  \bibfield  {author} {\bibinfo {author} {\bibfnamefont {D.}~\bibnamefont {Dosunmu}},\ }\href@noop {} {\enquote {\bibinfo {title} {When nigeria banned x, koo had a golden opportunity, and squandered it},}\ }\bibinfo {howpublished} {\url{https://restofworld.org/2023/nigeria-koo-twitter-rival-flop/} Accessed 10 November, 2023} (\bibinfo {year} {2023})\BibitemShut {NoStop}%
\bibitem [{\citenamefont {LiveMint}(2023)}]{livemint2023_koo}%
  \BibitemOpen
  \bibfield  {author} {\bibinfo {author} {\bibnamefont {LiveMint}},\ }\href@noop {} {\enquote {\bibinfo {title} {Not competing with twitter or threads, focusing on regional audiences: Koo co-founder},}\ }\bibinfo {howpublished} {\url{https://www.livemint.com/companies/not-competing-with-twitter-or-threads-focusing-on-regional-audiences-koo-cofounder-11692425534832.html} Accessed 10 November, 2023} (\bibinfo {year} {2023})\BibitemShut {NoStop}%
\bibitem [{\citenamefont {De~Choudhury}\ \emph {et~al.}(2010)\citenamefont {De~Choudhury}, \citenamefont {Sundaram}, \citenamefont {John}, \citenamefont {Seligmann},\ and\ \citenamefont {Kelliher}}]{choudhury2010_homophily}%
  \BibitemOpen
  \bibfield  {author} {\bibinfo {author} {\bibfnamefont {M.}~\bibnamefont {De~Choudhury}}, \bibinfo {author} {\bibfnamefont {H.}~\bibnamefont {Sundaram}}, \bibinfo {author} {\bibfnamefont {A.}~\bibnamefont {John}}, \bibinfo {author} {\bibfnamefont {D.~D.}\ \bibnamefont {Seligmann}}, \ and\ \bibinfo {author} {\bibfnamefont {A.}~\bibnamefont {Kelliher}},\ }\href {\doibase 10.48550/ARXIV.1006.1702} {\enquote {\bibinfo {title} {"birds of a feather": Does user homophily impact information diffusion in social media?}}\ } (\bibinfo {year} {2010})\BibitemShut {NoStop}%
\bibitem [{\citenamefont {Dehghani}\ \emph {et~al.}(2016)\citenamefont {Dehghani}, \citenamefont {Johnson}, \citenamefont {Hoover}, \citenamefont {Sagi}, \citenamefont {Garten}, \citenamefont {Parmar}, \citenamefont {Vaisey}, \citenamefont {Iliev},\ and\ \citenamefont {Graham}}]{Dehghani2016_homophily}%
  \BibitemOpen
  \bibfield  {author} {\bibinfo {author} {\bibfnamefont {M.}~\bibnamefont {Dehghani}}, \bibinfo {author} {\bibfnamefont {K.}~\bibnamefont {Johnson}}, \bibinfo {author} {\bibfnamefont {J.}~\bibnamefont {Hoover}}, \bibinfo {author} {\bibfnamefont {E.}~\bibnamefont {Sagi}}, \bibinfo {author} {\bibfnamefont {J.}~\bibnamefont {Garten}}, \bibinfo {author} {\bibfnamefont {N.~J.}\ \bibnamefont {Parmar}}, \bibinfo {author} {\bibfnamefont {S.}~\bibnamefont {Vaisey}}, \bibinfo {author} {\bibfnamefont {R.}~\bibnamefont {Iliev}}, \ and\ \bibinfo {author} {\bibfnamefont {J.}~\bibnamefont {Graham}},\ }\href {\doibase 10.1037/xge0000139} {\bibfield  {journal} {\bibinfo  {journal} {Journal of Experimental Psychology: General}\ }\textbf {\bibinfo {volume} {145}},\ \bibinfo {pages} {366} (\bibinfo {year} {2016})}\BibitemShut {NoStop}%
\bibitem [{\citenamefont {Al-garadi}\ \emph {et~al.}(2017)\citenamefont {Al-garadi}, \citenamefont {Varathan},\ and\ \citenamefont {Ravana}}]{Algaradi2017_kcore}%
  \BibitemOpen
  \bibfield  {author} {\bibinfo {author} {\bibfnamefont {M.~A.}\ \bibnamefont {Al-garadi}}, \bibinfo {author} {\bibfnamefont {K.~D.}\ \bibnamefont {Varathan}}, \ and\ \bibinfo {author} {\bibfnamefont {S.~D.}\ \bibnamefont {Ravana}},\ }\href {\doibase 10.1016/j.physa.2016.11.002} {\bibfield  {journal} {\bibinfo  {journal} {Physica A: Statistical Mechanics and its Applications}\ }\textbf {\bibinfo {volume} {468}},\ \bibinfo {pages} {278} (\bibinfo {year} {2017})}\BibitemShut {NoStop}%
\bibitem [{\citenamefont {Schiffman}(1998)}]{Schiffman1998_hindi}%
  \BibitemOpen
  \bibfield  {author} {\bibinfo {author} {\bibfnamefont {H.}~\bibnamefont {Schiffman}},\ }\href@noop {} {\emph {\bibinfo {title} {Linguistic culture and language policy}}},\ The Politics of Language\ (\bibinfo  {publisher} {Routledge},\ \bibinfo {address} {London, England},\ \bibinfo {year} {1998})\BibitemShut {NoStop}%
\bibitem [{\citenamefont {Kang}\ \emph {et~al.}(2023)\citenamefont {Kang}, \citenamefont {Xiao}, \citenamefont {Yu}, \citenamefont {Diaz},\ and\ \citenamefont {Zhang}}]{Kang2023_entropy}%
  \BibitemOpen
  \bibfield  {author} {\bibinfo {author} {\bibfnamefont {K.}~\bibnamefont {Kang}}, \bibinfo {author} {\bibfnamefont {Y.}~\bibnamefont {Xiao}}, \bibinfo {author} {\bibfnamefont {H.}~\bibnamefont {Yu}}, \bibinfo {author} {\bibfnamefont {M.~T.}\ \bibnamefont {Diaz}}, \ and\ \bibinfo {author} {\bibfnamefont {H.}~\bibnamefont {Zhang}},\ }\href {\doibase 10.3390/brainsci13111587} {\bibfield  {journal} {\bibinfo  {journal} {Brain Sciences}\ }\textbf {\bibinfo {volume} {13}},\ \bibinfo {pages} {1587} (\bibinfo {year} {2023})}\BibitemShut {NoStop}%
\bibitem [{\citenamefont {Kepinska}\ \emph {et~al.}(2023)\citenamefont {Kepinska}, \citenamefont {Caballero}, \citenamefont {Oliver}, \citenamefont {Marks}, \citenamefont {Haft}, \citenamefont {Zekelman}, \citenamefont {Kovelman}, \citenamefont {Uchikoshi},\ and\ \citenamefont {Hoeft}}]{Kepinska2023_entropy}%
  \BibitemOpen
  \bibfield  {author} {\bibinfo {author} {\bibfnamefont {O.}~\bibnamefont {Kepinska}}, \bibinfo {author} {\bibfnamefont {J.}~\bibnamefont {Caballero}}, \bibinfo {author} {\bibfnamefont {M.}~\bibnamefont {Oliver}}, \bibinfo {author} {\bibfnamefont {R.~A.}\ \bibnamefont {Marks}}, \bibinfo {author} {\bibfnamefont {S.~L.}\ \bibnamefont {Haft}}, \bibinfo {author} {\bibfnamefont {L.}~\bibnamefont {Zekelman}}, \bibinfo {author} {\bibfnamefont {I.}~\bibnamefont {Kovelman}}, \bibinfo {author} {\bibfnamefont {Y.}~\bibnamefont {Uchikoshi}}, \ and\ \bibinfo {author} {\bibfnamefont {F.}~\bibnamefont {Hoeft}},\ }\href {\doibase 10.1038/s41598-023-27952-2} {\bibfield  {journal} {\bibinfo  {journal} {Scientific Reports}\ }\textbf {\bibinfo {volume} {13}} (\bibinfo {year} {2023}),\ 10.1038/s41598-023-27952-2}\BibitemShut {NoStop}%
\bibitem [{\citenamefont {Newman}(2002)}]{Newman2002_assortativity}%
  \BibitemOpen
  \bibfield  {author} {\bibinfo {author} {\bibfnamefont {M.~E.~J.}\ \bibnamefont {Newman}},\ }\href {\doibase 10.1103/physrevlett.89.208701} {\bibfield  {journal} {\bibinfo  {journal} {Physical Review Letters}\ }\textbf {\bibinfo {volume} {89}} (\bibinfo {year} {2002}),\ 10.1103/physrevlett.89.208701}\BibitemShut {NoStop}%
\bibitem [{\citenamefont {Nagoshi}\ \emph {et~al.}(1990)\citenamefont {Nagoshi}, \citenamefont {Johnson},\ and\ \citenamefont {Danko}}]{Nagoshi1990_assortativity}%
  \BibitemOpen
  \bibfield  {author} {\bibinfo {author} {\bibfnamefont {C.~T.}\ \bibnamefont {Nagoshi}}, \bibinfo {author} {\bibfnamefont {R.~C.}\ \bibnamefont {Johnson}}, \ and\ \bibinfo {author} {\bibfnamefont {G.~P.}\ \bibnamefont {Danko}},\ }\href {\doibase 10.1007/bf01070737} {\bibfield  {journal} {\bibinfo  {journal} {Behavior Genetics}\ }\textbf {\bibinfo {volume} {20}},\ \bibinfo {pages} {23} (\bibinfo {year} {1990})}\BibitemShut {NoStop}%
\bibitem [{\citenamefont {Leszczensky}\ and\ \citenamefont {Pink}(2019)}]{Leszczensky2019_ethnicity}%
  \BibitemOpen
  \bibfield  {author} {\bibinfo {author} {\bibfnamefont {L.}~\bibnamefont {Leszczensky}}\ and\ \bibinfo {author} {\bibfnamefont {S.}~\bibnamefont {Pink}},\ }\href {\doibase 10.1177/0003122419846849} {\bibfield  {journal} {\bibinfo  {journal} {American Sociological Review}\ }\textbf {\bibinfo {volume} {84}},\ \bibinfo {pages} {394} (\bibinfo {year} {2019})}\BibitemShut {NoStop}%
\bibitem [{\citenamefont {Leonard}\ \emph {et~al.}(2007)\citenamefont {Leonard}, \citenamefont {Mehra},\ and\ \citenamefont {Katerberg}}]{Leonard2007_minority}%
  \BibitemOpen
  \bibfield  {author} {\bibinfo {author} {\bibfnamefont {A.~S.}\ \bibnamefont {Leonard}}, \bibinfo {author} {\bibfnamefont {A.}~\bibnamefont {Mehra}}, \ and\ \bibinfo {author} {\bibfnamefont {R.}~\bibnamefont {Katerberg}},\ }\href {\doibase 10.1002/job.488} {\bibfield  {journal} {\bibinfo  {journal} {Journal of Organizational Behavior}\ }\textbf {\bibinfo {volume} {29}},\ \bibinfo {pages} {573–589} (\bibinfo {year} {2007})}\BibitemShut {NoStop}%
\bibitem [{\citenamefont {Mehra}\ \emph {et~al.}(1998)\citenamefont {Mehra}, \citenamefont {Kilduff},\ and\ \citenamefont {Brass}}]{Mehra1998_minority}%
  \BibitemOpen
  \bibfield  {author} {\bibinfo {author} {\bibfnamefont {A.}~\bibnamefont {Mehra}}, \bibinfo {author} {\bibfnamefont {M.}~\bibnamefont {Kilduff}}, \ and\ \bibinfo {author} {\bibfnamefont {D.~J.}\ \bibnamefont {Brass}},\ }\href {\doibase 10.2307/257083} {\bibfield  {journal} {\bibinfo  {journal} {Academy of Management Journal}\ }\textbf {\bibinfo {volume} {41}},\ \bibinfo {pages} {441–452} (\bibinfo {year} {1998})}\BibitemShut {NoStop}%
\bibitem [{\citenamefont {Amato}\ \emph {et~al.}(2018)\citenamefont {Amato}, \citenamefont {Lacasa}, \citenamefont {Díaz-Guilera},\ and\ \citenamefont {Baronchelli}}]{Amato2018_language}%
  \BibitemOpen
  \bibfield  {author} {\bibinfo {author} {\bibfnamefont {R.}~\bibnamefont {Amato}}, \bibinfo {author} {\bibfnamefont {L.}~\bibnamefont {Lacasa}}, \bibinfo {author} {\bibfnamefont {A.}~\bibnamefont {Díaz-Guilera}}, \ and\ \bibinfo {author} {\bibfnamefont {A.}~\bibnamefont {Baronchelli}},\ }\href {\doibase 10.1073/pnas.1721059115} {\bibfield  {journal} {\bibinfo  {journal} {Proceedings of the National Academy of Sciences}\ }\textbf {\bibinfo {volume} {115}},\ \bibinfo {pages} {8260–8265} (\bibinfo {year} {2018})}\BibitemShut {NoStop}%
\bibitem [{\citenamefont {Scragg}(1974)}]{scragg1974_history}%
  \BibitemOpen
  \bibfield  {author} {\bibinfo {author} {\bibfnamefont {D.}~\bibnamefont {Scragg}},\ }\href {https://books.google.co.uk/books?id=oeNRAQAAIAAJ} {\emph {\bibinfo {title} {A History of English Spelling}}},\ Mont Follick series\ (\bibinfo  {publisher} {Manchester University Press},\ \bibinfo {year} {1974})\BibitemShut {NoStop}%
\bibitem [{\citenamefont {Vivian}(1979)}]{vivian1979_spelling}%
  \BibitemOpen
  \bibfield  {author} {\bibinfo {author} {\bibfnamefont {J.~H.}\ \bibnamefont {Vivian}},\ }\href@noop {} {\bibfield  {journal} {\bibinfo  {journal} {American Speech}\ }\textbf {\bibinfo {volume} {54}},\ \bibinfo {pages} {163} (\bibinfo {year} {1979})}\BibitemShut {NoStop}%
\bibitem [{\citenamefont {Abrams}\ and\ \citenamefont {Strogatz}(2003)}]{Abrams2003_language}%
  \BibitemOpen
  \bibfield  {author} {\bibinfo {author} {\bibfnamefont {D.~M.}\ \bibnamefont {Abrams}}\ and\ \bibinfo {author} {\bibfnamefont {S.~H.}\ \bibnamefont {Strogatz}},\ }\href {\doibase 10.1038/424900a} {\bibfield  {journal} {\bibinfo  {journal} {Nature}\ }\textbf {\bibinfo {volume} {424}},\ \bibinfo {pages} {900} (\bibinfo {year} {2003})}\BibitemShut {NoStop}%
\bibitem [{\citenamefont {Coleman}(1958)}]{coleman1958_homophily}%
  \BibitemOpen
  \bibfield  {author} {\bibinfo {author} {\bibfnamefont {J.~S.}\ \bibnamefont {Coleman}},\ }\href {http://www.jstor.org/stable/44124097} {\bibfield  {journal} {\bibinfo  {journal} {Human Organization}\ }\textbf {\bibinfo {volume} {17}},\ \bibinfo {pages} {28} (\bibinfo {year} {1958})}\BibitemShut {NoStop}%
\bibitem [{\citenamefont {Warton}\ \emph {et~al.}(2022)\citenamefont {Warton}, \citenamefont {Volny},\ and\ \citenamefont {Xu}}]{Warton2022_homophily}%
  \BibitemOpen
  \bibfield  {author} {\bibinfo {author} {\bibfnamefont {R.}~\bibnamefont {Warton}}, \bibinfo {author} {\bibfnamefont {C.}~\bibnamefont {Volny}}, \ and\ \bibinfo {author} {\bibfnamefont {K.~S.}\ \bibnamefont {Xu}},\ }in\ \href {\doibase 10.1007/978-3-031-17114-7_15} {\emph {\bibinfo {booktitle} {Social, Cultural, and Behavioral Modeling}}}\ (\bibinfo  {publisher} {Springer International Publishing},\ \bibinfo {year} {2022})\ pp.\ \bibinfo {pages} {155--164}\BibitemShut {NoStop}%
\bibitem [{\citenamefont {Are}\ \emph {et~al.}(2023)\citenamefont {Are}, \citenamefont {Card},\ and\ \citenamefont {Colijn}}]{Are2023_vaccine}%
  \BibitemOpen
  \bibfield  {author} {\bibinfo {author} {\bibfnamefont {E.~B.}\ \bibnamefont {Are}}, \bibinfo {author} {\bibfnamefont {K.~G.}\ \bibnamefont {Card}}, \ and\ \bibinfo {author} {\bibfnamefont {C.}~\bibnamefont {Colijn}},\ }\href {\doibase 10.1101/2023.06.06.23291056} {\  (\bibinfo {year} {2023}),\ 10.1101/2023.06.06.23291056}\BibitemShut {NoStop}%
\bibitem [{\citenamefont {Karat}(1972)}]{Karat1972_india}%
  \BibitemOpen
  \bibfield  {author} {\bibinfo {author} {\bibfnamefont {P.}~\bibnamefont {Karat}},\ }\href {\doibase 10.2307/3516456} {\bibfield  {journal} {\bibinfo  {journal} {Social Scientist}\ }\textbf {\bibinfo {volume} {1}},\ \bibinfo {pages} {25} (\bibinfo {year} {1972})}\BibitemShut {NoStop}%
\bibitem [{\citenamefont {Ronen}\ \emph {et~al.}(2014)\citenamefont {Ronen}, \citenamefont {Gonçalves}, \citenamefont {Hu}, \citenamefont {Vespignani}, \citenamefont {Pinker},\ and\ \citenamefont {Hidalgo}}]{ronen2014_gln}%
  \BibitemOpen
  \bibfield  {author} {\bibinfo {author} {\bibfnamefont {S.}~\bibnamefont {Ronen}}, \bibinfo {author} {\bibfnamefont {B.}~\bibnamefont {Gonçalves}}, \bibinfo {author} {\bibfnamefont {K.~Z.}\ \bibnamefont {Hu}}, \bibinfo {author} {\bibfnamefont {A.}~\bibnamefont {Vespignani}}, \bibinfo {author} {\bibfnamefont {S.}~\bibnamefont {Pinker}}, \ and\ \bibinfo {author} {\bibfnamefont {C.~A.}\ \bibnamefont {Hidalgo}},\ }\href {\doibase 10.1073/pnas.1410931111} {\bibfield  {journal} {\bibinfo  {journal} {Proceedings of the National Academy of Sciences}\ }\textbf {\bibinfo {volume} {111}},\ \bibinfo {pages} {E5616} (\bibinfo {year} {2014})},\ \Eprint {http://arxiv.org/abs/https://www.pnas.org/doi/pdf/10.1073/pnas.1410931111} {https://www.pnas.org/doi/pdf/10.1073/pnas.1410931111} \BibitemShut {NoStop}%
\bibitem [{\citenamefont {{Office of the Registrar General \& Census Commissioner, India (ORGI) }}(2011)}]{orgi2011_atlas}%
  \BibitemOpen
  \bibfield  {author} {\bibinfo {author} {\bibnamefont {{Office of the Registrar General \& Census Commissioner, India (ORGI) }}},\ }\href@noop {} {\enquote {\bibinfo {title} {Language atlas of india 2011},}\ }\bibinfo {howpublished} {\url{https://censusindia.gov.in/nada/index.php/catalog/42561} Accessed 5 February, 2024} (\bibinfo {year} {2011})\BibitemShut {NoStop}%
\bibitem [{\citenamefont {Yang}\ \emph {et~al.}(2021)\citenamefont {Yang}, \citenamefont {Choi}, \citenamefont {Abeliuk},\ and\ \citenamefont {Saffer}}]{Yang2021_discourse}%
  \BibitemOpen
  \bibfield  {author} {\bibinfo {author} {\bibfnamefont {A.}~\bibnamefont {Yang}}, \bibinfo {author} {\bibfnamefont {I.~M.}\ \bibnamefont {Choi}}, \bibinfo {author} {\bibfnamefont {A.}~\bibnamefont {Abeliuk}}, \ and\ \bibinfo {author} {\bibfnamefont {A.}~\bibnamefont {Saffer}},\ }\href {\doibase 10.1093/jcmc/zmab002} {\bibfield  {journal} {\bibinfo  {journal} {Journal of Computer-Mediated Communication}\ }\textbf {\bibinfo {volume} {26}},\ \bibinfo {pages} {148} (\bibinfo {year} {2021})}\BibitemShut {NoStop}%
\bibitem [{\citenamefont {Medina}\ \emph {et~al.}(2009)\citenamefont {Medina}, \citenamefont {Pokorny},\ and\ \citenamefont {Weigelt}}]{Medina2009_discourse}%
  \BibitemOpen
  \bibfield  {author} {\bibinfo {author} {\bibfnamefont {G.}~\bibnamefont {Medina}}, \bibinfo {author} {\bibfnamefont {B.}~\bibnamefont {Pokorny}}, \ and\ \bibinfo {author} {\bibfnamefont {J.}~\bibnamefont {Weigelt}},\ }\href {\doibase 10.1016/j.forpol.2008.11.004} {\bibfield  {journal} {\bibinfo  {journal} {Forest Policy and Economics}\ }\textbf {\bibinfo {volume} {11}},\ \bibinfo {pages} {392} (\bibinfo {year} {2009})}\BibitemShut {NoStop}%
\bibitem [{\citenamefont {Willis}(2019)}]{Willis2019_alpha}%
  \BibitemOpen
  \bibfield  {author} {\bibinfo {author} {\bibfnamefont {A.~D.}\ \bibnamefont {Willis}},\ }\href {\doibase 10.3389/fmicb.2019.02407} {\bibfield  {journal} {\bibinfo  {journal} {Frontiers in Microbiology}\ }\textbf {\bibinfo {volume} {10}} (\bibinfo {year} {2019}),\ 10.3389/fmicb.2019.02407}\BibitemShut {NoStop}%
\bibitem [{\citenamefont {Su}(2021)}]{Su2021_beta}%
  \BibitemOpen
  \bibfield  {author} {\bibinfo {author} {\bibfnamefont {X.}~\bibnamefont {Su}},\ }\href {\doibase 10.1128/msystems.00363-21} {\bibfield  {journal} {\bibinfo  {journal} {{mSystems}}\ }\textbf {\bibinfo {volume} {6}} (\bibinfo {year} {2021}),\ 10.1128/msystems.00363-21}\BibitemShut {NoStop}%
\bibitem [{\citenamefont {Grin}\ and\ \citenamefont {F\"{u}rst}(2022)}]{Grin2022_diversity}%
  \BibitemOpen
  \bibfield  {author} {\bibinfo {author} {\bibfnamefont {F.}~\bibnamefont {Grin}}\ and\ \bibinfo {author} {\bibfnamefont {G.}~\bibnamefont {F\"{u}rst}},\ }\href {\doibase 10.1007/s11205-022-02934-5} {\bibfield  {journal} {\bibinfo  {journal} {Social Indicators Research}\ }\textbf {\bibinfo {volume} {164}},\ \bibinfo {pages} {601} (\bibinfo {year} {2022})}\BibitemShut {NoStop}%
\bibitem [{\citenamefont {Mahfouz}(2020)}]{mahfouz2020_hashtags}%
  \BibitemOpen
  \bibfield  {author} {\bibinfo {author} {\bibfnamefont {I.~M.}\ \bibnamefont {Mahfouz}},\ }\href {\doibase 10.24093/awej/call6.6} {\bibfield  {journal} {\bibinfo  {journal} {Arab World English Journal}\ }\textbf {\bibinfo {volume} {6}},\ \bibinfo {pages} {84} (\bibinfo {year} {2020})}\BibitemShut {NoStop}%
\bibitem [{\citenamefont {Nguyen}\ \emph {et~al.}(2023)\citenamefont {Nguyen}, \citenamefont {Aljunied}, \citenamefont {Joty},\ and\ \citenamefont {Bing}}]{nguyen2023_llm}%
  \BibitemOpen
  \bibfield  {author} {\bibinfo {author} {\bibfnamefont {X.-P.}\ \bibnamefont {Nguyen}}, \bibinfo {author} {\bibfnamefont {S.~M.}\ \bibnamefont {Aljunied}}, \bibinfo {author} {\bibfnamefont {S.}~\bibnamefont {Joty}}, \ and\ \bibinfo {author} {\bibfnamefont {L.}~\bibnamefont {Bing}},\ }\href@noop {} {\enquote {\bibinfo {title} {Democratizing llms for low-resource languages by leveraging their english dominant abilities with linguistically-diverse prompts},}\ } (\bibinfo {year} {2023}),\ \Eprint {http://arxiv.org/abs/2306.11372} {arXiv:2306.11372 [cs.CL]} \BibitemShut {NoStop}%
\bibitem [{\citenamefont {Chiu}\ \emph {et~al.}(2014)\citenamefont {Chiu}, \citenamefont {Wang}, \citenamefont {Walther},\ and\ \citenamefont {Chao}}]{chu2014_chao}%
  \BibitemOpen
  \bibfield  {author} {\bibinfo {author} {\bibfnamefont {C.-H.}\ \bibnamefont {Chiu}}, \bibinfo {author} {\bibfnamefont {Y.-T.}\ \bibnamefont {Wang}}, \bibinfo {author} {\bibfnamefont {B.~A.}\ \bibnamefont {Walther}}, \ and\ \bibinfo {author} {\bibfnamefont {A.}~\bibnamefont {Chao}},\ }\href {\doibase https://doi.org/10.1111/biom.12200} {\bibfield  {journal} {\bibinfo  {journal} {Biometrics}\ }\textbf {\bibinfo {volume} {70}},\ \bibinfo {pages} {671} (\bibinfo {year} {2014})},\ \Eprint {http://arxiv.org/abs/https://onlinelibrary.wiley.com/doi/pdf/10.1111/biom.12200} {https://onlinelibrary.wiley.com/doi/pdf/10.1111/biom.12200} \BibitemShut {NoStop}%
\bibitem [{\citenamefont {Seshia}(1998)}]{Seshia1998_divide}%
  \BibitemOpen
  \bibfield  {author} {\bibinfo {author} {\bibfnamefont {S.}~\bibnamefont {Seshia}},\ }\href {\doibase 10.2307/2645684} {\bibfield  {journal} {\bibinfo  {journal} {Asian Survey}\ }\textbf {\bibinfo {volume} {38}},\ \bibinfo {pages} {1036–1050} (\bibinfo {year} {1998})}\BibitemShut {NoStop}%
\bibitem [{\citenamefont {Bhagat}(2001)}]{bhagat2001_divide}%
  \BibitemOpen
  \bibfield  {author} {\bibinfo {author} {\bibfnamefont {R.~B.}\ \bibnamefont {Bhagat}},\ }\href {http://www.jstor.org/stable/4411376} {\bibfield  {journal} {\bibinfo  {journal} {Economic and Political Weekly}\ }\textbf {\bibinfo {volume} {36}},\ \bibinfo {pages} {4352} (\bibinfo {year} {2001})}\BibitemShut {NoStop}%
\bibitem [{\citenamefont {Hao}\ \emph {et~al.}(2019)\citenamefont {Hao}, \citenamefont {Corral-Rivas}, \citenamefont {Gonz{\'{a}}lez-Elizondo}, \citenamefont {Ganeshaiah}, \citenamefont {Nava-Miranda}, \citenamefont {Zhang}, \citenamefont {Zhao},\ and\ \citenamefont {von Gadow}}]{Hao2019_bray}%
  \BibitemOpen
  \bibfield  {author} {\bibinfo {author} {\bibfnamefont {M.}~\bibnamefont {Hao}}, \bibinfo {author} {\bibfnamefont {J.~J.}\ \bibnamefont {Corral-Rivas}}, \bibinfo {author} {\bibfnamefont {M.~S.}\ \bibnamefont {Gonz{\'{a}}lez-Elizondo}}, \bibinfo {author} {\bibfnamefont {K.~N.}\ \bibnamefont {Ganeshaiah}}, \bibinfo {author} {\bibfnamefont {M.~G.}\ \bibnamefont {Nava-Miranda}}, \bibinfo {author} {\bibfnamefont {C.}~\bibnamefont {Zhang}}, \bibinfo {author} {\bibfnamefont {X.}~\bibnamefont {Zhao}}, \ and\ \bibinfo {author} {\bibfnamefont {K.}~\bibnamefont {von Gadow}},\ }\href {\doibase 10.1186/s40663-019-0188-9} {\bibfield  {journal} {\bibinfo  {journal} {Forest Ecosystems}\ }\textbf {\bibinfo {volume} {6}} (\bibinfo {year} {2019}),\ 10.1186/s40663-019-0188-9}\BibitemShut {NoStop}%
\bibitem [{\citenamefont {Emeneau}(1956)}]{emeneau1956_india}%
  \BibitemOpen
  \bibfield  {author} {\bibinfo {author} {\bibfnamefont {M.~B.}\ \bibnamefont {Emeneau}},\ }\href {http://www.jstor.org/stable/410649} {\bibfield  {journal} {\bibinfo  {journal} {Language}\ }\textbf {\bibinfo {volume} {32}},\ \bibinfo {pages} {3} (\bibinfo {year} {1956})}\BibitemShut {NoStop}%
\bibitem [{\citenamefont {Aula}(2014)}]{aula2014_language}%
  \BibitemOpen
  \bibfield  {author} {\bibinfo {author} {\bibfnamefont {S.}~\bibnamefont {Aula}},\ }\href@noop {} {\enquote {\bibinfo {title} {The problem with the english language in india},}\ }\bibinfo {howpublished} {\url{https://www.forbes.com/sites/realspin/2014/11/06/the-problem-with-the-english-language-in-india/} Accessed 12 November, 2023} (\bibinfo {year} {2014})\BibitemShut {NoStop}%
\bibitem [{\citenamefont {Chakravarty}(2023)}]{chakravarty2023_koo}%
  \BibitemOpen
  \bibfield  {author} {\bibinfo {author} {\bibfnamefont {S.}~\bibnamefont {Chakravarty}},\ }\href@noop {} {\enquote {\bibinfo {title} {Koo solves everything that annoys twitter users today, says koo ceo},}\ }\bibinfo {howpublished} {\url{https://www.outlookindia.com/business/koo-solves-everything-that-annoys-twitter-users-today-says-koo-ceo--news-256090} Accessed 29 September, 2023} (\bibinfo {year} {2023})\BibitemShut {NoStop}%
\bibitem [{\citenamefont {Cornish}(2022)}]{cornish2022_koo}%
  \BibitemOpen
  \bibfield  {author} {\bibinfo {author} {\bibfnamefont {C.}~\bibnamefont {Cornish}},\ }\href@noop {} {\enquote {\bibinfo {title} {India’s homegrown ‘twitter’ courts political parties to shed nationalist tag},}\ }\bibinfo {howpublished} {\url{https://www.ft.com/content/bd296702-18ad-4303-8192-bbb1268e3b56} Accessed 14 September, 2023} (\bibinfo {year} {2022})\BibitemShut {NoStop}%
\bibitem [{\citenamefont {Larsen}(2022)}]{larsen2022_sovereignty}%
  \BibitemOpen
  \bibfield  {author} {\bibinfo {author} {\bibfnamefont {B.~C.}\ \bibnamefont {Larsen}},\ }\href@noop {} {\enquote {\bibinfo {title} {The geopolitics of ai and the rise of digital sovereignty},}\ }\bibinfo {howpublished} {\url{https://www.brookings.edu/articles/the-geopolitics-of-ai-and-the-rise-of-digital-sovereignty/} Accessed 22 January, 2024} (\bibinfo {year} {2022})\BibitemShut {NoStop}%
\bibitem [{\citenamefont {Hershcovich}\ \emph {et~al.}(2022)\citenamefont {Hershcovich}, \citenamefont {Frank}, \citenamefont {Lent}, \citenamefont {de~Lhoneux}, \citenamefont {Abdou}, \citenamefont {Brandl}, \citenamefont {Bugliarello}, \citenamefont {Cabello~Piqueras}, \citenamefont {Chalkidis}, \citenamefont {Cui}, \citenamefont {Fierro}, \citenamefont {Margatina}, \citenamefont {Rust},\ and\ \citenamefont {Søgaard}}]{Hershcovich2022_nlp}%
  \BibitemOpen
  \bibfield  {author} {\bibinfo {author} {\bibfnamefont {D.}~\bibnamefont {Hershcovich}}, \bibinfo {author} {\bibfnamefont {S.}~\bibnamefont {Frank}}, \bibinfo {author} {\bibfnamefont {H.}~\bibnamefont {Lent}}, \bibinfo {author} {\bibfnamefont {M.}~\bibnamefont {de~Lhoneux}}, \bibinfo {author} {\bibfnamefont {M.}~\bibnamefont {Abdou}}, \bibinfo {author} {\bibfnamefont {S.}~\bibnamefont {Brandl}}, \bibinfo {author} {\bibfnamefont {E.}~\bibnamefont {Bugliarello}}, \bibinfo {author} {\bibfnamefont {L.}~\bibnamefont {Cabello~Piqueras}}, \bibinfo {author} {\bibfnamefont {I.}~\bibnamefont {Chalkidis}}, \bibinfo {author} {\bibfnamefont {R.}~\bibnamefont {Cui}}, \bibinfo {author} {\bibfnamefont {C.}~\bibnamefont {Fierro}}, \bibinfo {author} {\bibfnamefont {K.}~\bibnamefont {Margatina}}, \bibinfo {author} {\bibfnamefont {P.}~\bibnamefont {Rust}}, \ and\ \bibinfo {author} {\bibfnamefont {A.}~\bibnamefont {Søgaard}},\ }in\ \href {\doibase 10.18653/v1/2022.acl-long.482} {\emph {\bibinfo {booktitle} {Proceedings of the
  60th Annual Meeting of the Association for Computational Linguistics (Volume 1: Long Papers)}}}\ (\bibinfo  {publisher} {Association for Computational Linguistics},\ \bibinfo {year} {2022})\BibitemShut {NoStop}%
\end{thebibliography}%

\end{document}